# Discrimination in Online Ad Delivery

Latanya Sweeney
Harvard University
*latanya@fas.harvard.edu*

January 28, 2013[1]

## Abstract

A Google search for a person's name, such as "*Trevon Jones*", may yield a personalized ad for public records about Trevon that may be neutral, such as "*Looking for Trevon Jones? …*", or may be suggestive of an arrest record, such as "*Trevon Jones, Arrested?...*". This writing investigates the delivery of these kinds of ads by Google AdSense using a sample of racially associated names and finds statistically significant discrimination in ad delivery based on searches of 2184 racially associated personal names across two websites. First names, previously identified by others as being assigned at birth to more black or white babies, are found predictive of race (88% black, 96% white), and those assigned primarily to black babies, such as DeShawn, Darnell and Jermaine, generated ads suggestive of an arrest in 81 to 86 percent of name searches on one website and 92 to 95 percent on the other, while those assigned at birth primarily to whites, such as Geoffrey, Jill and Emma, generated more neutral copy: the word "arrest" appeared in 23 to 29 percent of name searches on one site and 0 to 60 percent on the other. On the more ad trafficked website, a black-identifying name was 25% more likely to get an ad suggestive of an arrest record. A few names did not follow these patterns: Dustin, a name predominantly given to white babies, generated an ad suggestive of arrest 81 and 100 percent of the time. All ads return results for actual individuals and ads appear regardless of whether the name has an arrest record in the company's database. Notwithstanding these findings, the company maintains Google received the same ad text for groups of last names (not first names), raising questions as to whether Google's advertising technology exposes racial bias in society and how ad and search technology can develop to assure racial fairness.

---

[1] v0.14 Preprint available at http://dataprivacylab.org/projects/onlineads/1071-1.pdf

## Introduction

*Have you ever been arrested*? Imagine the question not appearing in the solitude of your thoughts as you read this paper, but appearing explicitly whenever someone queries your name in a search engine.  Perhaps you are in competition for an award, an appointment, a promotion, or a new job, or maybe you are in a position of trust, such as a professor, a physician, a banker, a judge, a manager, or a volunteer, or perhaps you are completing a rental application, selling goods, applying for a loan, joining a social club, making new friends, dating, or engaged in any one of hundreds circumstances for which an online searcher seeks to learn more about you. Appearing alongside your list of accomplishments is an advertisement implying you may have a criminal record, whether you actually have one or not.  Worse, the ads don't appear for your competitors.

A person's criminal record begins when he is arrested for a crime. Job applications frequently include questions such as:

- "Have you ever been arrested?"
- "Have you ever been charged with a crime?"
- "Other than a traffic ticket, have you been convicted of a crime?"

Advantages of knowing such information when hiring or engaging with a person relate to trustworthiness.   Because others often equate a criminal record with not being reliable or honest, protections exist for those having criminal records.

If someone is falsely accused of a crime, pleads not guilty, and charges are dismissed, in the U.S., he may file suit against the person who brought the charges. For example, if a private citizen files a false criminal charge against you, or falsely makes a complaint to a police officer that results in your arrest, and if no conviction results, you may be able to sue the accuser for malicious prosecution.

If an employer disqualifies a job applicant based solely upon information indicating an arrest record, the company may face legal consequences. The U.S. Equal Employment Opportunity Commission ("EEOC") is the federal agency charged with enforcing Title VII of the Civil Rights Act of 1964, a law in the United States which applies to most employers, prohibiting employment discrimination based on race, color, religion, sex, or national origin, and through guidance issuance in 1973, extended to persons having criminal records [1,2].  Title VII does not prohibit employers from obtaining criminal background information.  However, certain uses of criminal information, such as a blanket policy or practice of excluding applicants or disqualifying employees based solely upon information indicating an arrest record, can result in a charge of discrimination. To make a determination, the EEOC uses an "adverse impact test," which measures whether practices, intentional or not, have a disproportionate effect. If the ratio of the effect on groups is less than 80%, the employer may be held responsible for discrimination [3].

So what about online ads suggesting someone with your name has an arrest record, even when no one with your name has ever been arrested?  The malicious prosecution approach does not apply.  Title VII does not apply either, unless you have an arrest record and can prove the potential employer used the ad or information from the company sponsoring the ad.

Further, the advertiser may argue that the ads are commercial free speech –a constitutional right to display the ad associated with your name.  The First Amendment of the U.S. Constitution protects advertising, as granted under the landmark U.S. Supreme Court decision, <u>Central Hudson Gas & Electric Corp. v. Public Service Commission of New York</u>, Supreme Court of the United States, 447 U.S. 557 (1980).  In <u>Central Hudson</u>, the Supreme Court sets out a four-part test for assessing government restrictions on commercial speech, which begins by determining whether the speech is misleading.  Are online ads suggesting the existence of an arrest record misleading if no one having the name has an arrest record?

Assume the ads are free speech: what happens when these ads appear more often for one racial group than another?  Not everyone is being equally affected by the free speech.  Is that free speech or is it racial discrimination?

*Racism* is "any attitude, action or institutional structure which subordinates a person or group because of their color . . . Racism is not just a matter of attitudes; actions and institutional structures can also be a form of racism" [4]. *Racial discrimination* results when a person or group of people is treated differently based on their racial origins [5]. Power is a necessary precondition, for it depends on the ability to give or withhold benefits, facilities, services, opportunities etc., from someone who should be entitled to them, and are denied on the basis of race. *Institutional or structural racism* is a system of procedures/patterns whose effect is to foster discriminatory outcomes or give preferences to members of one group over another [6].

Notice that racism can result, even if not intentional and that online activity may be so ubiquitous and intimately entwined with technology design that technologists may now have to think about societal consequences like structural racism in the technology they design. Such considerations are beyond this paper, but they frame the relevant legal, societal and technical landscape in which this work resides.

The investigation, chronicled in this writing, reports on an observed phenomenon, that some online ads suggestive of arrest records appear more often for one racial group than another among a sample of racially associated names. Because online ad delivery is a socio-technical construct, its study requires blending sociology and computer science, and so this writing presents such a blend.

## Problem Statement

> ***Given online searches of racially identifying names, show that associated personalized ads suggestive of an arrest record do not differ by race.***

Our hypothesis: no difference exists in the delivery of ads suggestive of an arrest record responding to online searches of racially associated names. Then, when presented with evidence of a pattern to the contrary, examine the pattern's credibility, likelihood and circumstances of occurring.

What is the suspected pattern of ad delivery? Below are three groups of ad hoc real-world examples that jointly describe concerns.

Earlier this year, a Google search for "*Latanya Farrell*" yielded the two ads appearing in Figure 1a. The first ad implies she may have been arrested, was she? After clicking on the link and paying the requisite subscription fee, we learn that the company has no arrest record for her (Figure 1b). A Google search for "*Latanya Sweeney*" and "*Latanya Lockett*" also yields ads suggestive of arrests. We find no arrest record for "*Latanya Sweeney*" but we do for "*Latanya Lockett*" (Figure 1). The ads appeared on [google.com](google.com) and on a newspaper website to which Google supplies ads, [reuters.com](reuters.com) (Figure 1c). All the ads in question link to instantcheckmate.com.

In comparison, searches for "*Kristen Haring*", "*Kristen Sparrow*" and "*Kristen Lindquist*" did not yield any instantcheckmate.com ads, only competitor ads (Figure 2a, 2c, and 2e), even though the company's database reports having records for all three names and arrest records for "*Kristen Sparrow*" and "*Kristen Lindquist*" (Figure 2d and 2f).

Searches for "*Jill Foley*", "*Jill Schneider*" and "*Jill James*" displayed instantcheckmate.com ads with neutral copy; the word "arrest" did not appear in the ads even though arrest records for all three names appear in the company's database (Figure 3).

Lastly, we consider a proxy for race associated with these names. Figure 4 shows Google images appearing for image searches of "*Latanya*", "*Latisha*", "*Kristen*" and "*Jill*", respectively. There appears a racial distinction. The faces associated with "*Latanya*" and "*Latisha*" (Figure 4a and 4b) tend to be black, while white faces dominate the images of "*Kristen*" and "*Jill*" (Figure 4c and 4d).

Together, these handpicked examples (Figures 2, 3 and 4) describe the suspected pattern –ads suggesting arrest tend to appear with names associated with blacks and neutral ads or no ads tend to appear with names associated with whites, regardless of whether the company has an arrest record associated with the name. The remainder of this paper describes a journey to establish an instance of the pattern worthy of scholarly consideration and statistical assessment.

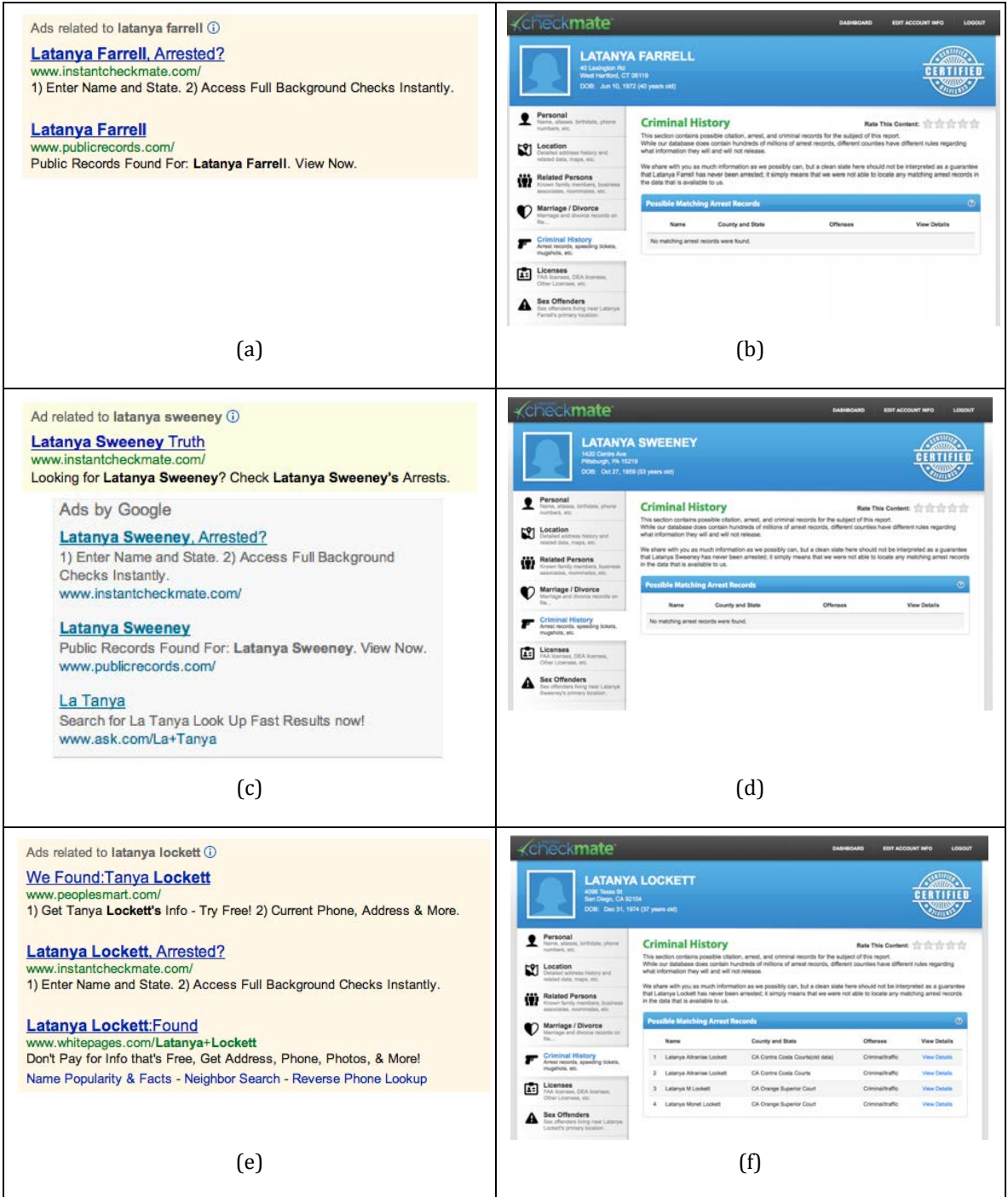

**Figure 1. Sample ads and criminal reports for "latanya farrell" (a,b), "latanya sweeney" (c,d), and "latanya locket"(e,f)  appearing on google.com (a,b,c) and reuters.com (c bottom). Criminal reports from instantcheckmate.com (b,d,f).**

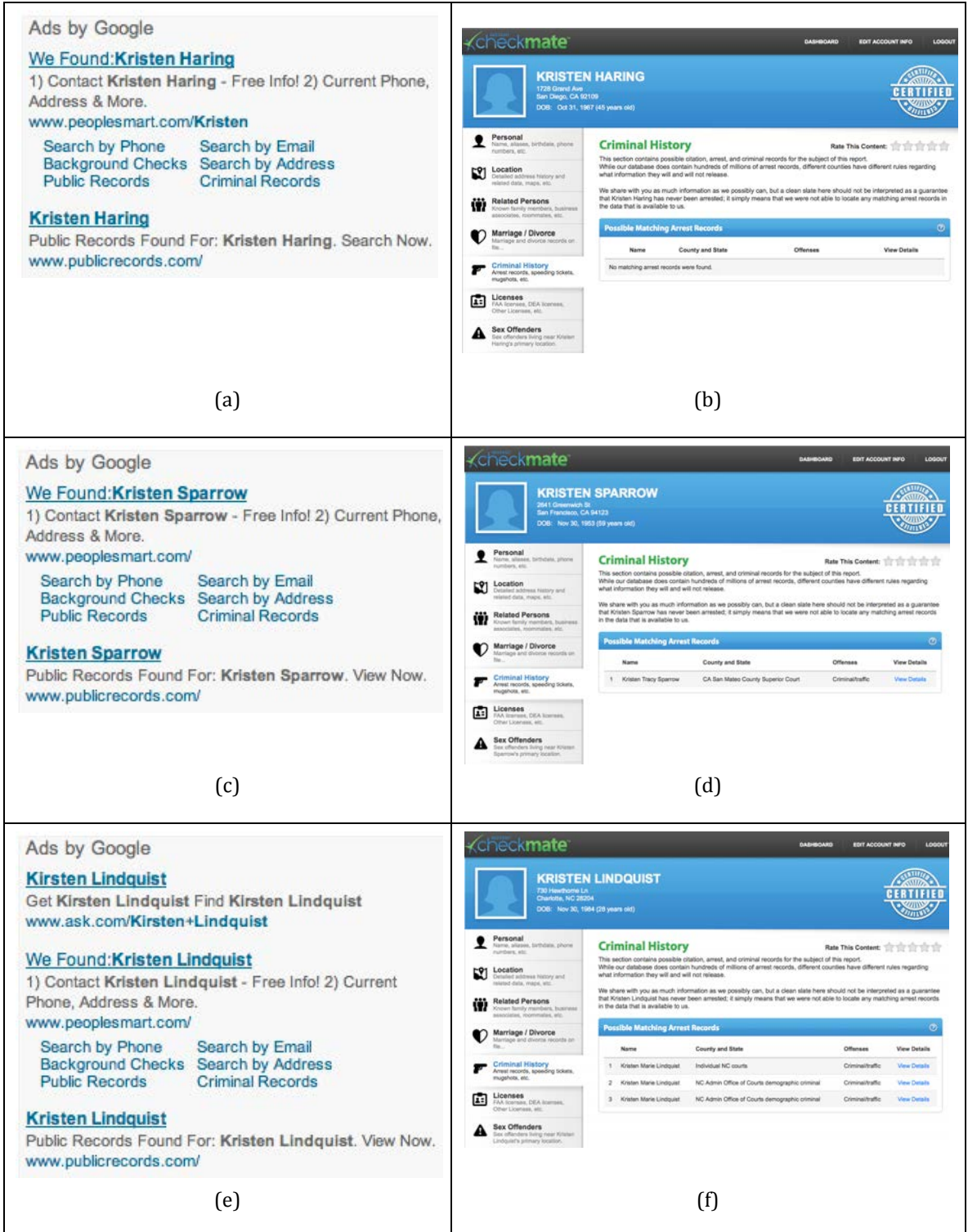

**Figure 2. Sample ads and criminal reports for "kristen haring" (a), "kristen sparrow" (b), and "kristen lindquist" (c), appearing on reuters.com (a,c,e). Criminal reports from instantcheckmate.com (b,d,f).**

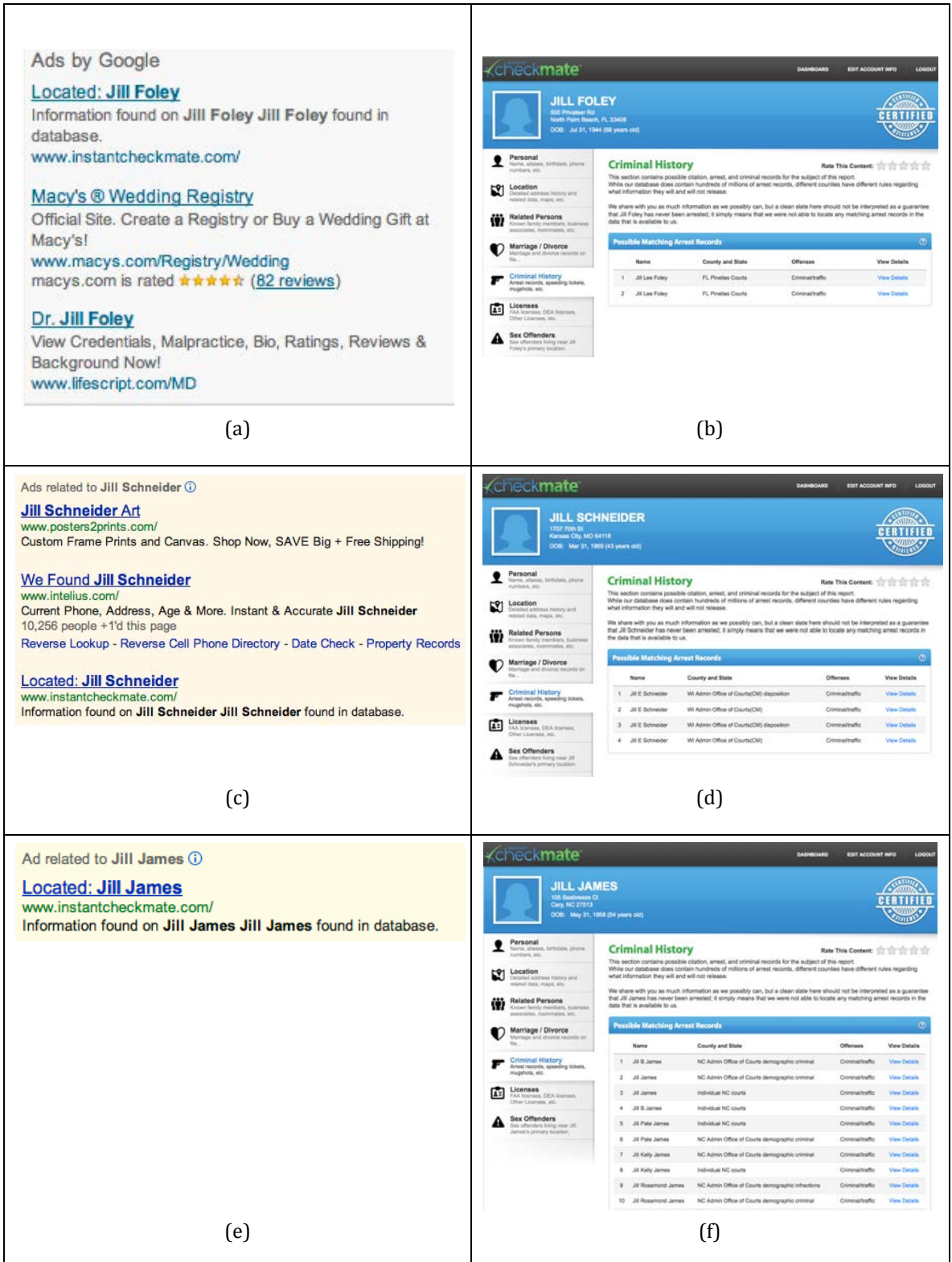

**Figure 3. Sample ads and criminal reports for "jill foley" (a,b), "jill schneider" (c,d), and "jill james"(e,f) appearing on google.com (c,e) and reuters.com (a). Criminal reports from instantcheckmate.com (b,d,f).**

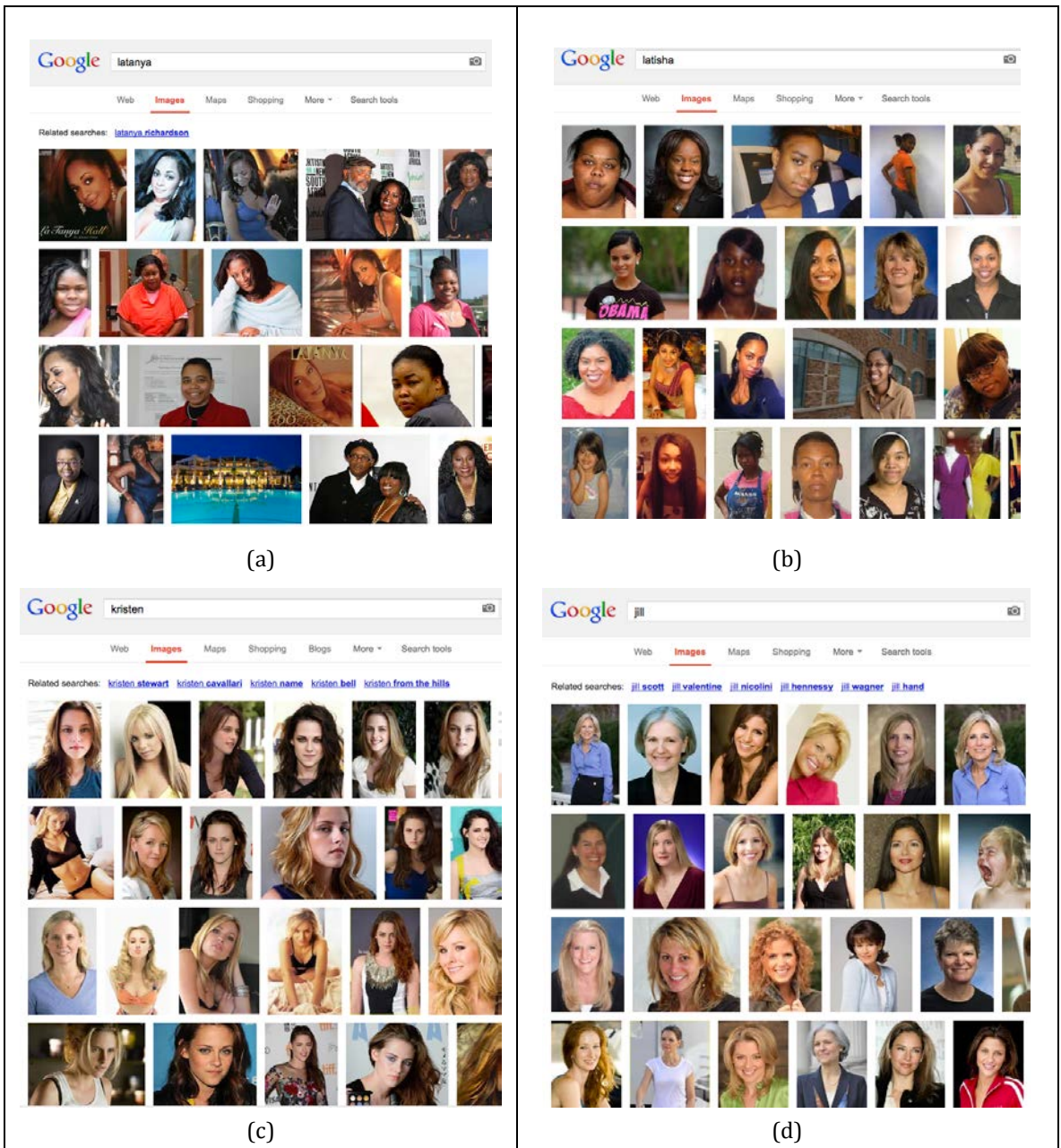

**Figure 4. Sample face images on google.com retrieved for searches "latanya" (a), "latisha" (b), "kristen" (c), and "jill" (d).**

## Google AdSense

Who generates the ad's text? Who decides when and where an ad will appear? What is the relationship between Google, Reuters and Instant Checkmate in the previous examples? An overview of Google AdSense, the program that delivered the ads in Figures 1, 2, and 3, explains entities and relationships.

In printed newspapers and magazines, ad space and ad content are fixed. Everyone who purchases the publication sees the same ad in the same space. But websites are different. Online ad space, not bound by the same physical limitations, can be dynamic, with ads tailored to the reader's search criteria, content interests, geographical location, and so on. Any two readers (or the same reader returning to the same website) might view different ads.

Google AdSense is the largest provider of dynamic online advertisements, placing ads for millions of sponsors on millions of websites [7]. In the first quarter of 2011, Google earned US $2.43 billion ($9.71 billion annualized), or 28% of total revenue, through Google AdSense [8]. AdSense has operational variations, but for simplicity, this writing only describes those features of Google AdSense specific to the Instant Checkmate ads in question.

When a reader enters search criteria in an enrolled website, Google AdSense embeds ads believed to be relevant to his search in the web page of results. Figures 1, 2, and 3 show ads delivered by Google AdSense in response to various "*firstname lastname*" searches.

To place an online ad, a "sponsor" provides Google with search criteria, copies of possible ads to deliver once a match occurs, and a financial bid (an amount the sponsor is willing to pay) if a reader clicks the delivered ad.[2] Google operates a real-time auction across bids for the same search criteria, usually displaying the ad having the highest bid first, the second highest second, and so on, and may elect not to show any ad if it considers the bid too low or if showing the ad exceeds a threshold (e.g. a maximum account total for the sponsor). In Figures 1, 2, and 3, Instant Checkmate sponsors the ads, which in most cases appears first among ads, implying Instant Checkmate had the highest bid.

A website owner wanting to "host" online ads enrolls in AdSense and changes his website to include special software that sends information about the current reader (e.g., search criteria) to Google and in exchange, receives corresponding ads from Google. The displayed ads have the banner "Ads by Google" when appearing on sites other than google.com. For example, reuters.com is an AdSense host, and entering "*Latanya Sweeney*" in the search bar at reuters.com generated a new web page having ads delivered by Google, bearing the banner "Ads by Google" (Figure 1c).

---

[2] This writing conflates two interacting Google programs: Google Adwords allows advertisers to specify search criteria, ad text and bids and Google AdSense delivers the ads to host sites.

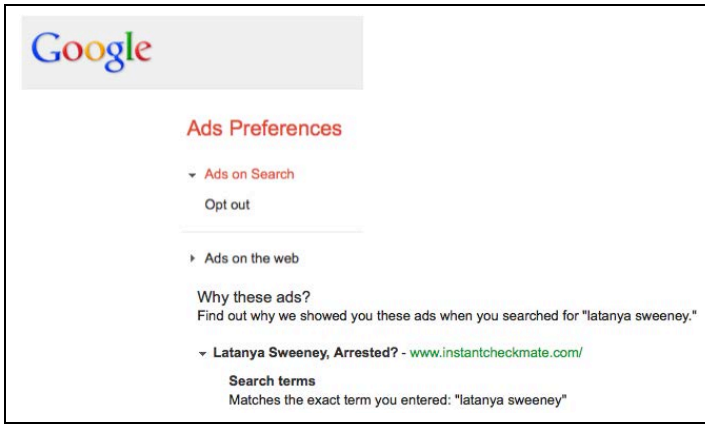

**Figure 5. Google explanation for delivering ad "*Latanya Sweeney, Arrested*?" –matches the exact first and last name searched.**

There is no cost associated with displaying an ad, but if the reader actually clicks the ad, the sponsor pays the promised bid, which is split between Google and the host. Clicking the "*Latanya Sweeney*" ad on reuters.com (Figure 1c) would cause Instant Checkmate to pay its bid to Google, which splits it with Reuters.

## Search Criteria

What search criteria did Instant Checkmate specify? Are ads randomly delivered? Do ads rely only on the first name?  Will ads be delivered for made-up names too? Google AdSense provides answers to these questions too.  Ads displayed on google.com allow readers to learn why a specific ad appeared. Clicking the circled "i" in the ad banner (e.g., Figure 1c) provides a web page explaining the ads (e.g., Figure 5).  Doing so for ads in Figures 1, 2, and 3, reveals that the ads appeared because the search criteria associated with the bid matched the exact first and last name combination searched.  Because bids presumably relate to records the company sells, the names would likely be the first and last names of real people, and because searches are online, ads may be more effective for people having online identities.

In summary, search criteria associated with ads:

- has to be both first and last names;
- should be names of real people; and,
- may prefer names of people with an online identity.

The next sections describe systematic construction of a list of racially associated first and last names for real people.  It is not presumed that Instant Checkmate placed bids or Google delivered ads using any such list. Instead, the list allows us to have a qualified sample of racially associated names for testing ad delivery.

**Black and White Identifying Names**

"Black-identifying" and "white-identifying" first names are those for which a significant number of people have the name and the frequency is sufficiently higher in one race than another. Heavily cited prior academic work provides exemplars.

In 2003, Bertrand and Mullainathan did a field experiment in which they provided resumes to job posts that were virtually identical except some of the resumes had black-identifying names and others had white-identifying names [9]. Their "Job Discrimination Study" showed significant discrimination against black names: white names received 50% more callbacks for interviews even though the resumes otherwise had identical qualifications.

The Job Discrimination study used a correlation of names given to black and white babies in Massachusetts between 1974 and 1979, defining black-identifying and white-identifying names as those that have the highest ratio of frequency in one racial group to frequency in the other racial group.

|     | White Female | Black Female | White Male | Black Male |
|-----|--------------|--------------|------------|------------|
| (a) | Allison      | Aisha        | Brad       | Darnell    |
|     | Anne         | Ebony        | Brendan    | Hakim      |
|     | Carrie       | Keisha       | Geoffrey   | Jermaine   |
|     | Emily        | Kenya        | Greg       | Kareem     |
|     | Jill         | Latonya      | Brett      | Jamal      |
|     | Laurie       | Lakisha      | Jay        | Leroy      |
|     | Kristen      | Latoya       | Matthew    | Rasheed    |
|     | Meredith     | Tamika       | Neil       | Tremayne   |
| (b) | Molly        | Imani        | Jake       | DeShawn    |
|     | Amy          | Ebony*       | Connor     | DeAndre    |
|     | Claire       | Shanice      | Tanner     | Marquis    |
|     | Emily*       | Aaliyah      | Wyatt      | Darnell*   |
|     | Katie        | Precious     | Cody       | Terrell    |
|     | Madeline     | Nia          | Dustin     | Malik      |
|     | Katelyn      | Deja         | Luke       | Trevon     |
|     | Emma         | Diamond      | Jack       | Tyrone     |
| (c) |              | Latanya      |            |            |
|     |              | Latisha      |            |            |

**Figure 6. Black-identifying and white-identifying first names from (a) the Job Discrimination Study [9], (b) Fryer and Levit [11], and (c) observation in Figure 4. Emily, a white female name, Ebony, a black female name, and Darnell, a black male name, appear in both (a) and (b), giving a total of 63 distinct first names.**

In the popular book "Freakonomics," Levitt and Dubner report the top 20 whitest- and blackest-identifying girl and boy names [10]. The list comes from earlier work by Fryer and Levitt, which shows a pattern change in the way Blacks named their children starting in the 1970's, which they correlate with the Black Power Movement [11]. They postulate that the movement influenced how Blacks perceived their identities and they give as evidence that before the movement, names given to black and white children were not distinctly different, but after the movement, the emergence of distinctly black names appear.

Similar to the Job Discrimination Study, the list used by Fryer and Levitt comes from names given to black and white children recorded in California birth records from 1961-2000 (over 16 million births).

We need a list of racially associated names in order to test ad delivery, so we use the union of lists from these prior studies augmented with two black female names, "*Latanya*" and "*Latisha*", from earlier observations. Figure 6 enumerates our list, having eight names for each of the categories: white female, black female, white male, and black male from the Job Discrimination Study (Figure 6a), and the first eight names for each category from the Fryer and Levitt work (Figure 6b). Removing duplicates gives a total of 63 distinct first names.

**Full Names of Real People**

Having a list of racially associated first names (Figure 6) is a start, but testing ad delivery requires a real person's first and last name ("full name"). How do we get full names? Web searches provide a means to locate and harvest full names by: (1) sampling names of professionals appearing on the Web; and, (2) sampling names of people active on social media sites and blogs ("netizens"). The subsections below describe the steps.

Harvesting Full Names of Professionals

Professionals often have their own websites or have biographical information appearing on institutional websites, listing titles and positions, and describing prior accomplishments and current activities. Several professions, such as research, medicine, law, and business, often have degree designations, such as PhD, MD, JD or MBA, associated with people in that profession. A Google search for a first name and a degree designation typically yields lists of people having that first name and degree. We use these kinds of searches to harvest a sample of full names of professionals having racially associated first names; Figure 8a itemizes the steps.

Here is a walk through the method of Figure 8a. The goal is to acquire a list of at least 10 full names for each racially associated first name. For each first name in the list of racially associated first names (Figure 6): perform a Google search with that first name and a degree designation (Step 1.1); harvest full names from the search results, up to 3 pages of results, avoiding duplicate names; and, for each full name recorded, visit its associated web page, and if an image is discernible, record whether the person appears black, white, or other. Archive each web page visited, preserving images and content.

Here are two examples. Figure 9a shows results for a Google search of "*Ebony PhD*". The results immediately reveal links for real people having "*Ebony*" as a first name – specifically, "*Ebony Bookman*", "*Ebony Glover*" (highlighted), "*Ebony Baylor*" and "*Ebony Utley*". We harvest the full names appearing on the first three pages of search results, using searches with other professional endings, such as *JD*, *MD*, or *MBA* as needed to find additional names in order to get at least 10 full names for "*Ebony*". Clicking on the link associated with "*Ebony Glover*" provides more information about her (Figure 9b), including an image. We record that the *Ebony Glover* in the study appears black.

Similarly, Figure 9c shows search results for "*Jill PhD*"—a list of professionals whose first name are *Jill*. Visiting links yields web pages with more information about each person. For example, Figure 9d shows an extract of *Jill Schneider* 's web page, and from the associated image, we record that the *Jill Schneider* in this study is white.

| Step 1 | **For each** *name$_i$* in the list of racially associated first names in Figure 6, **do**: |
|---|---|
| 1.1 | Perform a Google search for "*name$_i$ degree$_j$*" where *degree$_j$* is one of {*PhD, MD, JD, MBA*}. |
| 1.2 | **For each** result page, up to 3 pages, **do**: |
| | Preserve a copy of the page |
| | Record first and last names of people, avoiding duplicates. |
| | **For each** full name recorded, **do**: |
| | Click on the associated link. Preserve a copy of the resulting page. |
| | If personal image appears, record whether the person appears black, white, or other. |
| | **Repeat** Steps 1.1 and 1.2 with another *degree$_j$* if the number of full names for *name$_i$* is less than 10. |

(a)

| Step 1 | **For each** *name$_i$* in the list of racially associated first names in Figure 6, **do**: |
|---|---|
| 1.1 | Perform a PeekYou search for "*name$_i$*" |
| 1.2 | **For each** result page, up to 2 pages, and 10 recorded full names for *name$_i$* **do**: |
| | Preserve a copy of the page |
| | Record first and last names of people, avoiding duplicates. |
| | **For each** full name recorded, note whether associated image appears black, white, or other. |

(b)

**Figure 8. Method for harvesting racially associated first and last names of (a) professionals using Google search and (b) netizens using PeekYou.**

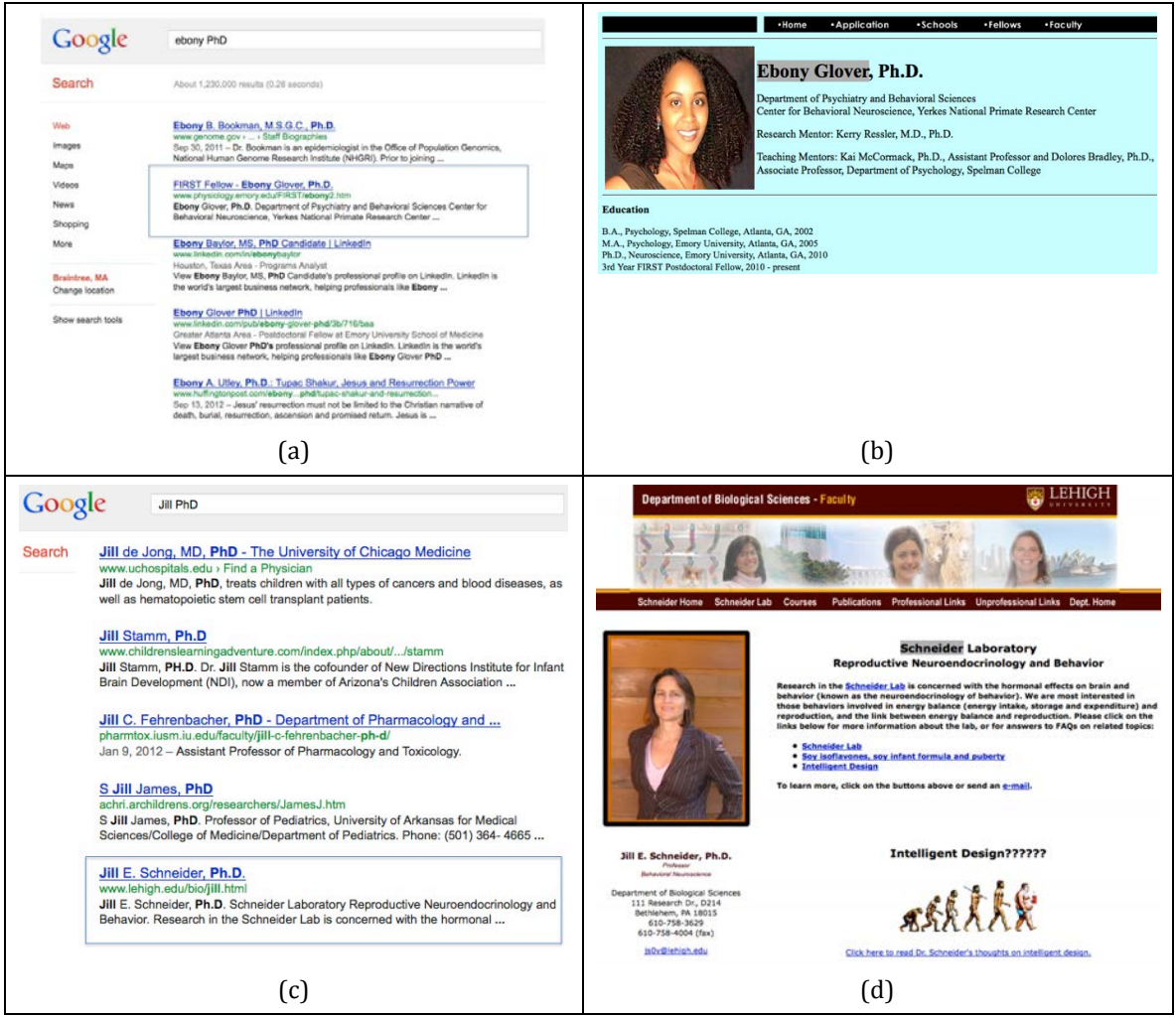

**Figure 9. Extracts of search and web pages for first names and degree designations. (a) Search "*Ebony Phd*". (b) "*Ebony Glover*" page. (c) Search "*Jill Phd*" (d) "*Jill Schneider*" page.**

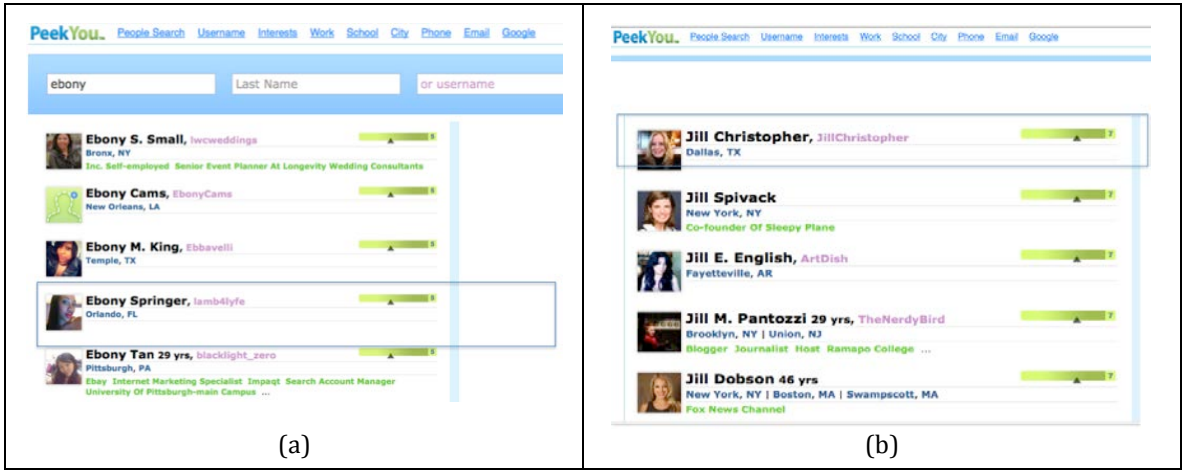

**Figure 10. Extracts of search pages for netizens using PeekYou.com for first names (a) "*Ebony*" and (b) "*Jill*". Highlighted records are (a) "*Ebony Springer*" and (b) "*Jill Christopher*".**

Harvesting Full Names of Netizens

The website peekyou.com ("PeekYou") compiles and disambiguates online and offline information on individuals, thereby connecting residential information with Facebook and twitter users, bloggers, and others, and assigns its own rating of size for each person's on-line footprint. Search results from peekyou.com ("PeekYou search") lists people having the highest score first, second highest second, and so on, and includes an image of the person. Celebrities and public figures tend to list first, having the highest PeekYou scores, followed by bloggers, tweeters and the rest. We use PeekYou searches to harvest a sample of full names of netizens having racially associated first names; Figure 8b itemizes the steps.

Harvesting names of netizens (Figure 8b) is similar but simpler than harvesting names of professionals (Figure 8a). For each name in the list of racially associated first names (Figure 6), perform a PeekYou search with that first name (Step 1.1); harvest full names from the search results, up to 2 pages of results, avoiding duplicate names; and, for each full name recorded, note whether the person in the associated image appears black, white, or other. Archive each web page, preserving images and content.

Here are two examples. Figure 10a shows some results from a PeekYou search of "*Ebony*" as a first name, listing "*Ebony Small*", "*Ebony Cams*", "*Ebony King*", "*Ebony Springer*" (highlighted), and "*Ebony Tan*". Similarly, Figure 10b shows some PeekYou search results for "*Jill*" as a first name, listing "*Jill Christopher*" (highlighted), "*Jill Spivack*", "*Jill English*", "*Jill Pantozzi*", and "*Jill Dobson*". We harvest these and other full names appearing on the first two pages of results and for each recorded image, report the race of the person if discernible. We record "*Ebony Glover*" in this study appears black and "*Jill Christopher*" white.

Results from Harvesting Full Names

Armed with the approach just described, from September 24 through October 22, 2012, I harvested 2184 racially associated full names of people with an online presence and using the images associated with those names, was able to confirm that the racially associated first names in Figure 6 are predictive of race (88% black and 96% white). Figures 11 and 12 summarize results. Below is a discussion.

Google searches of first names and degree designations were not as productive as first name lookups on PeekYou, 1002 to 1182 harvested names, respectively. White male names, "*Cody*", "*Connor*", "*Tanner*" and "*Wyatt*", retrieved results with those as last names not first names, the black male name, "*Kenya*", was confused with the country, and black names, "*Aaliyah*", "*Deja*", "*Diamond*", "*Hakim*", "*Malik*", "*Marquis*", "*Nia*", "*Precious*", "*Rasheed*" retrieved less than 10 full names. Only "*Diamond*" posed a problem with PeekYou searches –seemingly confused with other online entities. Other than "*Diamond*", all other searches contributed full names, and so unless noted otherwise, we exclude "*Diamond*" from further consideration.

| FIRST NAME | | | FULL NAMES | | | IMAGES | | | | | |
|---|---|---|---|---|---|---|---|---|---|---|---|
| Name | Race | Gender | Professionals | Netizens | Full Names | None | Black | % | White | % | Other |
| Aaliyah | Black | Female | 5 | 14 | 19 | 7 | 12 | 100% | 0 | 0% | 0 |
| Aisha | Black | Female | 22 | 32 | 54 | 13 | 36 | 88% | 5 | 12% | 0 |
| Allison | White | Female | 14 | 14 | 28 | 4 | 0 | 0% | 23 | 96% | 1 |
| Amy | White | Female | 42 | 25 | 67 | 24 | 0 | 0% | 40 | 93% | 3 |
| Anne | White | Female | 19 | 16 | 35 | 7 | 1 | 4% | 27 | 96% | 0 |
| Brad | White | Male | 19 | 18 | 37 | 13 | 0 | 0% | 24 | 100% | 0 |
| Brendan | White | Male | 15 | 25 | 40 | 16 | 3 | 13% | 20 | 83% | 1 |
| Brett | White | Male | 13 | 15 | 28 | 5 | 0 | 0% | 23 | 100% | 0 |
| Carrie | White | Female | 17 | 16 | 33 | 10 | 0 | 0% | 21 | 91% | 2 |
| Claire | White | Female | 30 | 26 | 56 | 18 | 1 | 3% | 37 | 97% | 0 |
| Cody | White | Male | 0 | 30 | 30 | 14 | 0 | 0% | 16 | 100% | 0 |
| Connor | White | Male | 0 | 30 | 30 | 6 | 1 | 4% | 23 | 96% | 0 |
| Darnell | Black | Male | 12 | 14 | 26 | 11 | 14 | 93% | 1 | 7% | 0 |
| DeAndre | Black | Male | 16 | 13 | 29 | 14 | 15 | 100% | 0 | 0% | 0 |
| Deja | Black | Female | 0 | 24 | 24 | 3 | 18 | 86% | 3 | 14% | 0 |
| DeShawn | Black | Male | 13 | 14 | 27 | 14 | 11 | 85% | 2 | 15% | 0 |
| Diamond | Black | Female | 0 | 0 | 0 | | | | | | |
| Dustin | White | Male | 36 | 30 | 66 | 24 | 0 | 0% | 42 | 100% | 0 |
| Ebony | Black | Female | 14 | 45 | 59 | 9 | 46 | 92% | 4 | 8% | 0 |
| Emily | White | Female | 15 | 15 | 30 | 3 | 0 | 0% | 26 | 96% | 1 |
| Emma | White | Female | 33 | 27 | 60 | 27 | 2 | 6% | 30 | 91% | 1 |
| Geoffrey | White | Male | 19 | 15 | 34 | 4 | 0 | 0% | 29 | 97% | 1 |
| Greg | White | Male | 20 | 20 | 40 | 17 | 0 | 0% | 23 | 100% | 0 |
| Hakim | Black | Male | 4 | 13 | 17 | 9 | 7 | 88% | 1 | 13% | 0 |
| Imani | Black | Female | 12 | 13 | 25 | 6 | 19 | 100% | 0 | 0% | 0 |
| Jack | White | Male | 28 | 30 | 58 | 30 | 0 | 0% | 24 | 86% | 4 |
| Jake | White | Male | 29 | 30 | 59 | 27 | 1 | 3% | 30 | 94% | 1 |
| Jamal | Black | Male | 11 | 18 | 29 | 8 | 10 | 48% | 4 | 19% | 7 |
| Jay | White | Male | 15 | 14 | 29 | 6 | 3 | 13% | 18 | 78% | 2 |
| Jermaine | Black | Male | 14 | 14 | 28 | 13 | 15 | 100% | 0 | 0% | 0 |
| Jill | White | Female | 20 | 14 | 34 | 4 | 0 | 0% | 30 | 100% | 0 |
| Kareem | Black | Male | 18 | 15 | 33 | 10 | 17 | 74% | 0 | 0% | 6 |
| Katelyn | White | Female | 50 | 30 | 80 | 29 | 0 | 0% | 51 | 100% | 0 |
| Katie | White | Female | 37 | 13 | 50 | 26 | 0 | 0% | 24 | 100% | 0 |
| Keisha | Black | Female | 11 | 29 | 40 | 11 | 28 | 97% | 1 | 3% | 0 |
| Kenya | Black | Female | 4 | 0 | 4 | 1 | 3 | 100% | 0 | 0% | 0 |
| Kristen | White | Female | 20 | 14 | 34 | 7 | 0 | 0% | 27 | 100% | 0 |
| Lakisha | Black | Female | 13 | 15 | 28 | 15 | 13 | 100% | 0 | 0% | 0 |
| Latanya | Black | Female | 13 | 15 | 28 | 13 | 13 | 87% | 2 | 13% | 0 |
| Latisha | Black | Female | 13 | 15 | 28 | 7 | 19 | 90% | 2 | 10% | 0 |
| Latonya | Black | Female | 21 | 15 | 36 | 11 | 23 | 92% | 2 | 8% | 0 |
| Latoya | Black | Female | 12 | 15 | 27 | 12 | 15 | 100% | 0 | 0% | 0 |
| Laurie | White | Female | 13 | 15 | 28 | 7 | 1 | 5% | 20 | 95% | 0 |
| Leroy | Black | Male | 11 | 14 | 25 | 7 | 9 | 50% | 7 | 39% | 2 |
| Luke | White | Male | 35 | 25 | 60 | 31 | 1 | 3% | 27 | 93% | 1 |
| Madeline | White | Female | 37 | 29 | 66 | 36 | 2 | 7% | 28 | 93% | 0 |
| Malik | Black | Male | 1 | 17 | 18 | 6 | 12 | 100% | 0 | 0% | 0 |
| Marquis | Black | Male | 5 | 14 | 19 | 3 | 15 | 94% | 1 | 6% | 0 |
| Matthew | White | Male | 18 | 26 | 44 | 12 | 0 | 0% | 32 | 100% | 0 |
| Meredith | White | Female | 18 | 15 | 33 | 5 | 2 | 7% | 26 | 93% | 0 |
| Molly | White | Female | 41 | 29 | 70 | 27 | 0 | 0% | 42 | 98% | 1 |
| Neil | White | Male | 18 | 12 | 30 | 11 | 0 | 0% | 17 | 89% | 2 |
| Nia | Black | Female | 0 | 11 | 11 | 0 | 8 | 73% | 3 | 27% | 0 |
| Precious | Black | Female | 0 | 12 | 12 | 4 | 7 | 88% | 1 | 13% | 0 |
| Rasheed | Black | Male | 1 | 16 | 17 | 6 | 8 | 73% | 2 | 18% | 1 |
| Shanice | Black | Female | 12 | 14 | 26 | 6 | 18 | 90% | 1 | 5% | 1 |
| Tamika | Black | Female | 14 | 15 | 29 | 12 | 17 | 100% | 0 | 0% | 0 |
| Tanner | White | Male | 0 | 30 | 30 | 12 | 0 | 0% | 18 | 100% | 0 |
| Terrell | Black | Male | 13 | 15 | 28 | 7 | 17 | 81% | 4 | 19% | 0 |
| Tremayne | Black | Male | 15 | 12 | 27 | 14 | 12 | 92% | 1 | 8% | 0 |
| Trevon | Black | Male | 12 | 14 | 26 | 12 | 14 | 100% | 0 | 0% | 0 |
| Tyrone | Black | Male | 19 | 17 | 36 | 13 | 19 | 83% | 3 | 13% | 1 |
| Wyatt | White | Male | 0 | 30 | 30 | 17 | 0 | 0% | 13 | 100% | 0 |
| | | Totals | 1002 | 1182 | 2184 | 756 | 508 | | 881 | | 39 |

**Figure 11. Summary of harvesting 2184 full names of professionals and neitzens from the Web (middle group) using racially associated first names (leftmost group), and race observations of online images (rightmost group). A total of 1428 images, 508 black, 881 white and 39 other.**

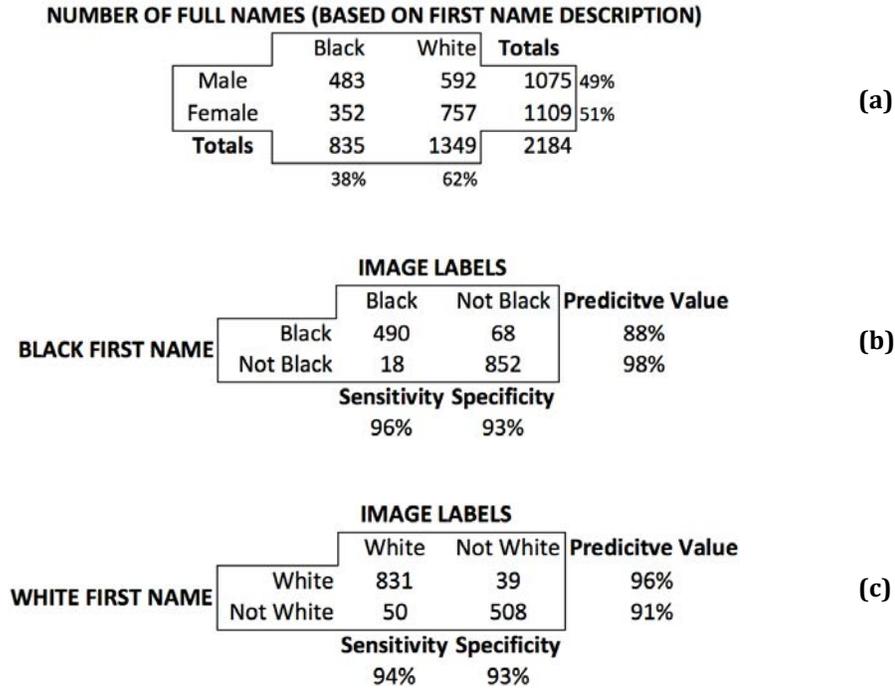

**NUMBER OF FULL NAMES (BASED ON FIRST NAME DESCRIPTION)**

|        | Black | White | Totals |     |
|--------|-------|-------|--------|-----|
| Male   | 483   | 592   | 1075   | 49% |
| Female | 352   | 757   | 1109   | 51% |
| **Totals** | 835 | 1349 | 2184 |     |
|        | 38%   | 62%   |        |     |

(a)

**IMAGE LABELS**

**BLACK FIRST NAME**

|           | Black | Not Black | Predicitve Value |
|-----------|-------|-----------|------------------|
| Black     | 490   | 68        | 88%              |
| Not Black | 18    | 852       | 98%              |
|           | Sensitivity | Specificity |          |
|           | 96%   | 93%       |                  |

(b)

**IMAGE LABELS**

**WHITE FIRST NAME**

|           | White | Not White | Predicitve Value |
|-----------|-------|-----------|------------------|
| White     | 831   | 39        | 96%              |
| Not White | 50    | 508       | 91%              |
|           | Sensitivity | Specificity |          |
|           | 94%   | 93%       |                  |

(c)

**Figure 12. Descriptive statistics of harvested full names (a) and analysis of first names as a classifier for blacks (b) and for whites (c).**

Figure 11 shows the number of full names harvested for each first name. Names contributing the most number of full names have white first names, e.g. "*Katelyn*" (80), "*Molly*" (70), "*Amy*" (67), "*Dustin*" (66) and "*Madeline*" (66), purposefully oversampled to test whether comparable PeekYou scores have any effect on ad delivery. Names contributing the least number of full names have black first names, "*Hakim*" (17), "*Rasheed*" (17), "*Precious*" (12), "*Nia*" (11) and "*Kenya*" (4).

The average number of full names for each first name is 35, with a median of 30, and standard deviation 16. For black first names, the average number of full names for each of the 31 first names is 27, with median 27, and standard deviation 11, and for the 31 white first names, the average is 44, median 35, and standard deviation 16.

Of the 2184 full names harvested, 835 (38%) are associated with black first names and 1349 (62%) with white first names, and 1075 (49%) with male first names and 1109 (51%) with female names; see Figure 12a.

Most images associated with black-identifying names were of black people (88%) and an even greater percentage of images associated with white-identifying names were of white people (96%). A total of 1428 names had discernible black (508), white (881) or other (39) images (Figure 11). We examine black and white names separately as predictors of race (Figures 12b and 12c). Of those having black associated first names, 490 images were of blacks, 68 images were not, 18 images having white first names were of blacks, and 852 names had neither black first names nor images of blacks.  Similarly, 831 images of whites had white first names, 50 images of whites did not have white first names, 39 had white first names but non-white images, and 508 had neither white first names nor images of whites.

Some first names associated as black had perfect predictions (100%) –"*Aaliyah*", "*DeAndre*", "*Imani*", "*Jermaine*", "*Lakisha*", "*Latoya*", "*Malik*", "*Tamika*", and "*Trevon*" —and the worst predictors of blacks were "*Jamal*" (48%) and "*Leroy*" (50%).  Figure 11 has details. Even more first names associated with whites, 12 of 31 names or 39%, made perfect predictions –"*Brad*", "*Brett*", "*Cody*", "*Dustin*", "*Greg*", "*Jill*", "*Katelyn*", "*Katie*", "*Kristen*", "*Matthew*", "*Tanner*" and "*Wyatt*" –and the worst predictors of whites, "*Jay*" (78%) and "*Brendan*" (83%"), were not bad. These findings strongly support the use of these names as racial indicators in this study.

Sixty-two full names (or 62/2184 = 3%) appeared in the list twice even though the people were not necessarily the same.   No name appeared more than twice, so overall, Google and PeekYou searches tended to yield different names.

## Ad Delivery

We now have a set of first and last names suggestive of race.  What ads appear when these names are searched? To answer this question, we examine ads delivered on two sites, Google.com and Reuters.com, in response to searches of each full name, once at each site.

The method is straightforward.  For each full name, visit Google.com, search for the name and record which ads display.  Repeat the process at Reuters.com, clearing the browser's cache and cookies before each search and preserving copies of web pages received.  Figure 13 enumerates these steps.

As examples, Figure 14 shows ads delivered in response to searches of "*Lakisha Simmons*", "*Laurie Ryan*", "*Darnell Bacon*", and "*Brendan Watson*" on google.com and reuters.com.  We preserve the capture of all ads, not just those of Instant Checkmate.

| Step 1 | **For each** *fullname*ᵢ in the list of racially associated full names, **do**: |
|---|---|
| 1.1 | Clear the browser cache and cookies. |
| 1.2 | Search Google.com for "*fullname*ᵢ" |
| 1.3 | Preserve copies of any of up to the first 3 pages of results having ads. |
| 1.4 | Record which ads display. |
| | |
| 1.5 | Clear the browser cache and cookies. |
| 1.6 | Search Reuters.com for "*fullname*ᵢ" |
| 1.7 | Preserve a copy of the resulting page. |
| 1.8 | Record which ads display. |

**Figure 13. Method for harvesting ads appearing in responses to searches of full names on google.com and reuters.com.**

**(a)** Ad related to Lakisha Simmons ⓘ
**Lakisha simmons: Truth**
www.instantcheckmate.com/
Arrests and Much More. Everything About Lakisha simmons

**(b)** Ads by Google
**Lakisha Simmons, Arrested?**
1) Enter Name and State. 2) Access Full Background Checks Instantly.
www.instantcheckmate.com/

**We Found:Lakisha Simmons**
1) Contact Lakisha Simmons - Free Info! 2) Current Phone, Address & More.
www.peoplesmart.com/Lakisha
   Search by Phone      Search by Email
   Background Checks   Search by Address
   Public Records      Criminal Records

**We Found Lakeisha Simmons**
Current Address, Phone and Age. Find Lakeisha simmons, Anywhere.
www.peoplefinders.com/

**(c)** Ad related to Laurie Ryan ⓘ
**We Found Laurie Ryan**
www.whitepages.com/Laurie+Ryan
Get Phone, Address & More for Laurie Ryan, Try Free Now!
Name Popularity & Facts - Neighbor Search - Reverse Phone Lookup

**(d)** Ads by Google
**Background Of Laurie Ryan**
Search Instant Checkmate The Records Of Laurie Ryan
www.instantcheckmate.com/

**We Found:Ryan Laurie**
1) Get Ryan Laurie's Background Report 2) Contact Info & More - Try Free!
www.peoplesmart.com/
   Search by Phone      Search by Email
   Background Checks   Search by Address
   Public Records      Criminal Records

**Laurie Ryan**
Public Records Found For: Laurie Ryan. View Now.
www.publicrecords.com/

**(e)** Ad related to Darnell Bacon ⓘ
**Darnell Bacon, Arrested?**
www.instantcheckmate.com/
1) Enter Name and State. 2) Access Full Background Checks Instantly.

**(f)** Ads by Google
**Darnell Bacon, Arrested?**
1) Enter Name and State. 2) Access Full Background Checks Instantly.
www.instantcheckmate.com/

**Darnell Bacon**
Public Records Found For: Darnell Bacon. View Now.
www.publicrecords.com/
   People Search       Public Records Search
   Background Check    Criminal Check

**We Found:Darnell Bacon**
1) Contact Darnell Bacon - Free Info! 2) Current Phone, Address & More.
www.peoplesmart.com/Darnell
   Search by Phone      Search by Email
   Background Checks   Search by Address
   Public Records      Criminal Records

**(g)** Ad related to Brendan Watson ⓘ
**Records For Brendan Watson**
www.instantcheckmate.com/
View Anyone's Criminal History. Check Criminal Records In Seconds!

**(h)** Ads by Google
**Located: Brendan Watson**
Information found on Brendan Watson Brendan Watson found in database.
www.instantcheckmate.com/

**We Found:Brenden Watson**
1) Get Brenden's Background Report 2) Contact Info & More - Try Free!
www.peoplesmart.com/
   Search by Phone      Search by Email
   Background Checks   Search by Address
   Public Records      Criminal Records

**Brenden Watson**
Public Records Found For: Brenden Watson. View Now.
www.publicrecords.com/

**Figure 14. Ads in response to full name searches on google.com (a,c,e,g) and reuters.com (b,d,f,h) for "*Lakisha Simmons*", "*Laurie Ryan*", "*Darnell Bacon*", and "*Brendan Watson*".**

Results from Ad Delivery

From September 24 through October 23, 2012, I searched 2184 full names on google.com and reuters.com, as described above. Execution took place at different times of days, different days of week, with different IP and machine addresses operating in different parts of the United States using different browsers. I manually searched 1373 of the names and used automated means ("Webshot" [12]) for the remainder (812 names). Here are 15 findings about ads and names, followed by four supplemental observations.

1. No more than three ads ever appeared for a search, whether manual or automated, regardless of website, Google or Reuters. No company's ad listed more than once on a page.

2. Far fewer ads appeared on google.com than on reuters.com. A total of 5337 ads appeared, 4473 (84%) on reuters.com and only 864 (16%) on Google, even when examining up to three pages of search results on google.com, and Google showed fewer ads per page, typically 1 (median) compared to 3 (median) on Reuters. In terms of the 2184 full names, ads appeared exclusive to Reuters (1221), Google (17) and on both (604) for a total of 1842 (84%) names having ads; 342 names had no ads at all. Reuters displayed ads for 1826 (84%) names and Google for 622 (28%). Figures 15a and 15c have summary statistics.

3. Most ads were for government-collected information ("public records") about the person. Public records in the United States often include a person's address, phone number, criminal history, and professional and business licenses, though specifics vary among states. Of the 5337 total ads captured, all but 1161 were for public records, or conversely, 4176 ads (78% of all ads) were for public records. Figure 15a has a distribution.

4. Ads for public records appeared for most names. Of the 2184 names, 1705 (78%) had at least one ad for public records about the person being searched. Reuters showed ads for 1598 names and Google for 544 names. Figure 16 has details.

5. More Instant Checkmate ads appeared than for any other company. Four companies accounted for more than half of all ads: instantcheckmate.com (1557 of 5337 or 29%), publicrecords.com (861 or 16%), peoplesmart.com (589, 11%), and peoplefinders.com (542, 10%). All ads for these companies sold public records. Ad distribution was different on Google's site; Instant Checkmate still had the most ads (431 of 864 or 50%), but Intelius, another seller of public records, while not in the top four overall, had the second most ads (127 or 15%) on google.com. Figure 15a lists details.

6. Instant Checkmate ads dominated the topmost ad position. On reuters.com, ads for Instant Checkmate listed first in 892 (49%) of the 1826 searches having ads on Reuters. The next closest, publicrecords.com, was a distance back having the topmost spot only 142 times, but most frequently appearing in the second and third positions. Figure 15b summarizes ad positions.

7. Ads for public records appeared more often in black names than white. Regardless of company, proportionately more ads appeared for names having a black-identifying first name. PeopleSmart ads appeared for 270 white and 280 black names, being disproportionately higher for blacks, 41% (280 of 679) to 29%. PublicRecords ads appeared 10% more often for black (54%) than white (44%) names, and Instant Checkmate ads 2.45% more often for blacks (72% to 69%). Figure 15d lists findings.

8. Instant Checkmate ads accounted for the largest percentage of ads in most first name categories, except for "*Kristen*", "*Connor*", and "*Tremayne*", which have uncharacteristically fewer ads. Instant Checkmate ads appeared for an average of 70% of all full names in a first name group receiving ads on Reuters (median 76%, standard deviation 0.21, 63 first name groups). For example, Instant Checkmate ads appeared on Reuters for 90-100% of all full names having ads whose name began "*Kenya*", "*Latoya*", "*DeShawn*", "*Emily*", "*Jay*", "*Greg*", "*Brendan*", "*Brad*", "*Leroy*", "*Dustin*", "*Neil*" or "*Jill*". In three cases, Instant Checkmate ads fell under 25% despite competition: "*Tremayne*" (91% PublicRecords, 23% Instant Checkmate), "*Connor*" (80% PublicRecords, 20% Instant Checkmate), and "*Kristen*" (58% PublicRecords, 16% Instant Checkmate). Figure 16 shows results by first name group.

9. Instant Checkmate had the most variability in ad copy. Almost all ads for public records included the name of the person in the ad itself, making each ad virtually unique, but beyond personalization, there was little variability in ad templates. Of the 534 PeopleFinder ads appearing on Reuters, all but 11 used the same personalized template, "We found *fullname.* Current Address, Phone and Age. Find *fullname*, Anywhere", where the person's first and last name replaces *fullname*. PublicRecords used 5 templates and PeopleSmart 7, but Instant Checkmate used more than all others combined, 18 templates in 1126 ads. Figure 17 displays ad texts and frequencies for all four companies.

10. Only Instant Checkmate ads included the word "arrest". While Instant Checkmate's competitors, PeopleSmart, PublicRecords, and PeopleFinders, also sell criminal history information, none of their ads included the word "arrest". In the 18 templates of Instant Checkmate ads found on Reuters, 8 of them include the word "arrest"; see Figure 17 for details.

11. Instant Checkmate ads having "arrest" in its text appeared less often than ads not including the word on Reuters. Of the 1126 Instant Checkmate ads appearing on Reuters, 544 (48%) include the word "arrest" and 582 (52%) do not. Figure 19 provides details.

12. A greater percentage of Instant Checkmate ads having "arrest" in ad text appeared for black identifying first names than for white first names. Of the 1126 Instant Checkmate ads on Reuters, 488 displayed with black-identifying first names, 291 (60%) of which had "arrest" in ad text. Of the 638 ads displayed with white-identifying names, 308 (48%) had "arrest". These results are statistically significant, $x^2(1)$=14.32, $p < 0.001$; there is less than a 0.1% probability that these data can be explained by chance. The results also have an adverse impact ratio (40%/52%) of 77%, satisfying the EEOC's and U.S. Department of Labor's 80% adverse impact test if this were employment. Figure 15e shows analysis.

13. More white identifying first names top the list of neutral Instant Checkmate Ads than do black names. On reuters.com, the highest percentage of neutral ads, where the word "arrest" does not appear in ad text, were ads for "*Jill*" (77%) and "*Emma*" (75%), both white-identifying names. Names receiving the highest percentage of ads with "arrest" in the text are "*Darnell*" (84%), "*Jermaine*" (81%) and "*DeShawn*" (86%), all black-identifying first names. Some names appear opposite this pattern. "*Dustin*", a white-identifying name, generated "arrest" ads in 81% of searches with that first name, and "*Imani*", a black-identifying name, received neutral copy in 75% of "*Imani*" searches. Figure 19 provides results by first name groups.

14. Instant Checkmate ads appearing on google.com often used different ad text than on Reuters. While the same neutral and arrest ads having dominant appearances on Reuters also appeared frequently on Google, ads on google.com included an additional 10 templates, all using the word "criminal", a word also suggestive of arrest, or the word "arrest". These new templates appeared in 89 of the 432 ads (21%). Figure 20 lists the Instant Checkmate ad templates found on google.com.

15. On google.com, a greater percentage of Instant Checkmate ads suggestive of arrest displayed for black associated first names than white. Of the 432 Instant Checkmate ads appearing on google.com, 90% (388) were suggestive of arrest regardless of race. Of the 366 ads that appeared for black-identifying names, 335 (92%) were suggestive of arrest. Far fewer ads displayed for white-identifying names (66 total), and 53 (80%) were suggestive of arrest. These results are statistically significant, $x^2(1)$=7.71, $p < 0.01$; there is less than a 1% probability that these data can be explained by chance. The adverse impact ratio (8%/20%) of 40%, which would satisfy the EEOC adverse impact test if this were employment. Figure 15f shows analysis and Figures 21 and 22 show distributions.

Here are four supplemental observations.

16. A greater percentage of Instant Checkmate ads having the word "arrest" in ad text appeared for black identifying first names than for white identifying first names within professional and netizen subsets. Of the 2184 names in the study, 599, harvested using professional designations, had Instant Checkmate ads on Reuters with 217 having black associated names, 136 (63%) of which received ads with the word "arrest" in ad text compared to only 178 (47%) of 382 white associated names, a statistically significant difference ($X^2(1)=14.34, p < 0.001$). Netizens also had a higher percentage of black names having ads with the word "arrest" in 155 (57%) of 271 ads for black identifying names compared to 130 (51%) of 256 ads for white identifying names.

17. People behind the names used in this study are diverse. Examining source webpages for the names reveals all kinds of people. Political figures include State Representatives Aisha Braveboy ("arrest" ad) and Jay Jacobs (neutral ad) of Maryland, Jill Biden (neutral ad), wife of U.S. Vice President Joe Biden, and Claire McCaskill, whose campaign advertisement for the U.S. Senate is alongside an Instant Checkmate ad having the word "arrest" (Figure 23). Names mined from academic websites include graduate students, researchers, administrators, staff, and accomplished academics, such as Amy Gutmann, President of the University of Pennsylvania and Chair of the U.S. Presidential Commission for the Study of Bioethical Issues. Dustin Hoffman ("arrest" ad) is among names of celebrities. A smorgasbord of athletes appears, from local to national fame, including numerous high school stars (assorted neutral and "arrest" ads). The youngest person associated with the study was a missing 11-year-old black girl.

18. PeekYou, the primary source of names for Netizens in this study, assigns a score to each name estimating the name's overall presence on the Web. As expected, celebrities get the highest scores, 10's and 9's. Of the 2184 names in the study, 1143 were harvested from PeekYou with scores, and only 4 of these had a PeekYou score of 10 and 12 had a 9 score. Dustin Hoffman is a 9. Only 2 ads appeared for these high scoring names. Other than that, an abundance of ads appeared across the remaining spectrum of PeekYou scores. Figure 25 shows distributions of Peek You scores.

19. Different Instant Checkmate ads appear for the same person. Of the 2184 names, 228 names had Instant Checkmate ads on both Reuters and Google, but only 42 of these received the same ad. The other 186 (82%) names received different ads across the two sites. Search results on Reuters for the 62 duplicate names that appeared in the study show different ads for 37 (60%) names, the same ad for 7 names, and no ad for 18. At most, three distinct ads appeared across Reuters and Google for the same name; Figure 24 has examples.

|  | Reuters | Google | Totals |  |
|---|---|---|---|---|
| instantcheckmate | 1126 | 431 | 1557 | 29% |
| peoplesmart | 550 | 39 | 589 | 11% |
| publicrecords | 770 | 91 | 861 | 16% |
| peoplefinders | 535 | 7 | 542 | 10% |
| facebook | 29 |  | 29 | 1% |
| intelius | 56 | 127 | 183 | 3% |
| whitepages | 61 | 33 | 94 | 2% |
| ask | 175 | 13 | 188 | 4% |
| usa-people-search | 73 |  | 73 | 1% |
| other (public records) | 53 | 7 | 60 | 1% |
| not public records | 1045 | 116 | 1161 | 22% |
| Totals | 4473 | 864 | 5337 |  |
|  | 84% | 16% |  |  |

(a)

**Ad Position on Reuters**

|  | 1st | 2nd | 3rd | Totals |
|---|---|---|---|---|
| instantcheckmate | **892** | 157 | 77 | 1126 |
| peoplesmart | 110 | 252 | 188 | 550 |
| publicrecords | 142 | **350** | **278** | 770 |
| peoplefinders | 128 | 202 | 205 | 535 |
| facebook | 2 | 9 | 18 | 29 |
| intelius | 5 | 25 | 26 | 56 |
| whitepages | 7 | 17 | 37 | 61 |
| ask | 60 | 73 | 42 | 175 |
| usa-people-search | 12 | 27 | 34 | 73 |
| other (public records) | 14 | 19 | 20 | 53 |

|  |  |
|---|---|
| Ads for public records | 3428 |
| Other ads (not public records) | 1045 |
| Total ads on Reuters | 4473 |

(b)

|  | Reuters | Google | Totals |
|---|---|---|---|
| **Names having Ads** | 1826 | 622 |  |
| Exclusive to Reuters or Google | 1221 | 17 | 1238 |
| Ads/Name: Average | 2.4 | 1.4 |  |
| Ads/Name: Median | 3 | 1 |  |
| Standard Deviation | 0.8 | 0.7 |  |
| **Names having No Ads** |  |  | 342 |
| **Names with Ads on Both Reuters and Google** |  |  | 604 |
| Total Full Names |  |  | 2184 |

(c)

**Public Record Ads on Reuters by Type of First Name**

|  | BLACK |  | WHITE |  | Totals | Ratio | Difference (p-value) |
|---|---|---|---|---|---|---|---|
| instantcheckmate | 488 | 72% | 638 | 69% | 1126 | 1.04 | 2.45% |
| peoplesmart | 280 | 41% | 270 | 29% | 550 | 1.40 | 11.86% |
| publicrecords | 368 | 54% | 402 | 44% | 770 | 1.24 | 10.45% |
| peoplefinders | 264 | 39% | 271 | 29% | 535 | 1.32 | 9.39% |
| intelius | 37 | 5% | 19 | 2% | 56 |  | 3.38% |
| Total names | 679 |  | 919 |  | 1598 |  |  |

(d)

**INSTANT CHECKMATE ADS ON REUTERS**

|  | OBSERVED |  |  |  | EXPECTED |  |
|---|---|---|---|---|---|---|
|  | BLACK | WHITE | Totals |  | BLACK | WHITE |
| Arrest Ads | 291  60% | 308  48% | 599  53% |  | 260 | 339 |
| Neutral Ads | 197  40% | 330  52% | 527  47% |  | 228 | 299 |
| Totals | 488 | 638 | 1126 |  |  |  |

(e)

**INSTANT CHECKMATE ADS ON GOOGLE**

|  | OBSERVED |  |  |  | EXPECTED |  |
|---|---|---|---|---|---|---|
|  | BLACK | WHITE | Totals |  | BLACK | WHITE |
| Arrest Ads | 335  92% | 53  80% | 388  90% |  | 329 | 59 |
| Neutral Ads | 31  8% | 13  20% | 44  10% |  | 37 | 7 |
| Totals | 366 | 66 | 432 |  |  |  |

(f)

**Figure 15.** Summary statistics for (a) ads appearing on Reuters and Google; (b) ad positions on Reuters; (c) results by names; (d) ads for public record appearing on Reuters by racially associated first name; (e) Chi-Square test for Instant Checkmate ads on Reuters; and, (f) Chi-Square test for Instant Checkmate ads on Google.

| FIRST NAME | | | NAMES WITH PUBLIC RECORD ADS | | | NUMBER OF FULL NAMES WITH PUBLIC RECORD ADS ON REUTERS | | | | | | | | | |
|---|---|---|---|---|---|---|---|---|---|---|---|---|---|---|---|
| Name | Race | Full Names | Reuters | Google | Distinct | instantcheckmate | | peoplesmart | | publicrecords | | peoplefinders | | intelius | |
| Aaliyah | Black | 19 | 13 | 3 | 14 | 4 | 31% | 2 | 15% | 13 | 100% | 0 | 0% | 8 | 62% |
| Aisha | Black | 54 | 47 | 26 | 50 | 37 | 79% | 24 | 51% | 25 | 53% | 26 | 55% | 1 | 2% |
| Allison | White | 28 | 17 | 0 | 19 | 12 | 71% | 2 | 12% | 6 | 35% | 7 | 41% | 1 | 6% |
| Amy | White | 67 | 41 | 0 | 43 | 27 | 66% | 18 | 44% | 19 | 46% | 10 | 24% | 0 | 0% |
| Anne | White | 35 | 20 | 2 | 20 | 16 | 80% | 0 | 0% | 6 | 30% | 5 | 25% | 2 | 10% |
| Brad | White | 37 | 32 | 9 | 32 | 30 | 94% | 13 | 41% | 9 | 28% | 4 | 13% | 0 | 0% |
| Brendan | White | 40 | 36 | 12 | 36 | 34 | 94% | 10 | 28% | 15 | 42% | 6 | 17% | 0 | 0% |
| Brett | White | 28 | 25 | 4 | 25 | 21 | 84% | 6 | 24% | 9 | 36% | 4 | 16% | 0 | 0% |
| Carrie | White | 33 | 23 | 7 | 24 | 17 | 74% | 6 | 26% | 7 | 30% | 1 | 4% | 0 | 0% |
| Claire | White | 56 | 40 | 0 | 43 | 31 | 78% | 14 | 35% | 7 | 18% | 13 | 33% | 0 | 0% |
| Cody | White | 30 | 16 | 0 | 17 | 9 | 56% | 8 | 50% | 13 | 81% | 1 | 6% | 0 | 0% |
| Connor | White | 30 | 20 | 0 | 21 | 4 | 20% | 2 | 10% | 16 | 80% | 1 | 5% | 1 | 5% |
| Darnell | Black | 26 | 23 | 20 | 25 | 19 | 83% | 6 | 26% | 12 | 52% | 7 | 30% | 0 | 0% |
| DeAndre | Black | 29 | 21 | 9 | 24 | 15 | 71% | 6 | 29% | 13 | 62% | 8 | 38% | 3 | 14% |
| Deja | Black | 24 | 15 | 13 | 19 | 11 | 73% | 0 | 0% | 11 | 73% | 7 | 47% | 1 | 7% |
| DeShawn | Black | 27 | 22 | 13 | 22 | 21 | 95% | 7 | 32% | 16 | 73% | 1 | 5% | 1 | 5% |
| Diamond | Black | | | | | | | | | | | | | | |
| Dustin | White | 66 | 52 | 4 | 53 | 47 | 90% | 20 | 38% | 28 | 54% | 6 | 12% | 0 | 0% |
| Ebony | Black | 59 | 47 | 28 | 49 | 39 | 83% | 11 | 23% | 28 | 60% | 15 | 32% | 0 | 0% |
| Emily | White | 30 | 20 | 0 | 21 | 19 | 95% | 2 | 10% | 4 | 20% | 1 | 5% | 0 | 0% |
| Emma | White | 60 | 36 | 0 | 39 | 20 | 56% | 17 | 47% | 22 | 61% | 8 | 22% | 1 | 3% |
| Geoffrey | White | 34 | 27 | 7 | 29 | 24 | 89% | 9 | 33% | 13 | 48% | 5 | 19% | 0 | 0% |
| Greg | White | 40 | 37 | 11 | 39 | 35 | 95% | 6 | 16% | 10 | 27% | 14 | 38% | 1 | 3% |
| Hakim | Black | 17 | 9 | 4 | 11 | 5 | 56% | 2 | 22% | 4 | 44% | 5 | 56% | 1 | 11% |
| Imani | Black | 25 | 16 | 8 | 18 | 8 | 50% | 4 | 25% | 11 | 69% | 7 | 44% | 4 | 25% |
| Jack | White | 58 | 31 | 0 | 35 | 25 | 81% | 9 | 29% | 13 | 42% | 10 | 32% | 0 | 0% |
| Jake | White | 59 | 29 | 1 | 32 | 20 | 69% | 8 | 28% | 8 | 28% | 8 | 28% | 2 | 7% |
| Jamal | Black | 29 | 24 | 11 | 26 | 11 | 46% | 9 | 38% | 8 | 33% | 16 | 67% | 1 | 4% |
| Jay | White | 29 | 19 | 5 | 20 | 18 | 95% | 2 | 11% | 3 | 16% | 5 | 26% | 0 | 0% |
| Jermaine | Black | 28 | 22 | 20 | 22 | 16 | 73% | 14 | 64% | 13 | 59% | 7 | 32% | 0 | 0% |
| Jill | White | 34 | 29 | 12 | 29 | 26 | 90% | 7 | 24% | 11 | 38% | 8 | 28% | 0 | 0% |
| Kareem | Black | 33 | 28 | 19 | 28 | 21 | 75% | 11 | 39% | 13 | 46% | 6 | 21% | 1 | 4% |
| Katelyn | White | 80 | 63 | 0 | 72 | 26 | 41% | 8 | 13% | 34 | 54% | 54 | 86% | 3 | 5% |
| Katie | White | 50 | 26 | 0 | 27 | 14 | 54% | 13 | 50% | 10 | 38% | 7 | 27% | 1 | 4% |
| Keisha | Black | 40 | 30 | 14 | 36 | 24 | 80% | 16 | 53% | 8 | 27% | 13 | 43% | 3 | 10% |
| Kenya | Black | 4 | 4 | 2 | 4 | 4 | 100% | 1 | 25% | 0 | 0% | 3 | 75% | 2 | 50% |
| Kristen | White | 34 | 19 | 1 | 22 | 3 | 16% | 10 | 53% | 11 | 58% | 7 | 37% | 0 | 0% |
| Lakisha | Black | 28 | 26 | 18 | 26 | 15 | 58% | 15 | 58% | 5 | 19% | 17 | 65% | 0 | 0% |
| Latanya | Black | 28 | 24 | 17 | 24 | 19 | 79% | 19 | 79% | 13 | 54% | 4 | 17% | 0 | 0% |
| Latisha | Black | 28 | 22 | 19 | 23 | 17 | 77% | 5 | 23% | 12 | 55% | 20 | 91% | 0 | 0% |
| Latonya | Black | 36 | 34 | 17 | 34 | 26 | 76% | 17 | 50% | 17 | 50% | 14 | 41% | 1 | 3% |
| Latoya | Black | 27 | 26 | 20 | 27 | 25 | 96% | 9 | 35% | 15 | 58% | 6 | 23% | 0 | 0% |
| Laurie | White | 28 | 19 | 6 | 19 | 14 | 74% | 5 | 26% | 8 | 42% | 6 | 32% | 0 | 0% |
| Leroy | Black | 25 | 24 | 22 | 24 | 22 | 92% | 13 | 54% | 11 | 46% | 5 | 21% | 0 | 0% |
| Luke | White | 60 | 38 | 0 | 38 | 23 | 61% | 18 | 47% | 18 | 47% | 4 | 11% | 2 | 5% |
| Madeline | White | 66 | 41 | 0 | 53 | 11 | 27% | 17 | 41% | 24 | 59% | 36 | 88% | 1 | 2% |
| Malik | Black | 18 | 14 | 8 | 15 | 8 | 57% | 2 | 14% | 4 | 29% | 7 | 50% | 3 | 21% |
| Marquis | Black | 19 | 17 | 13 | 17 | 14 | 82% | 5 | 29% | 9 | 53% | 0 | 0% | 0 | 0% |
| Matthew | White | 44 | 29 | 9 | 29 | 24 | 83% | 7 | 17% | 10 | 34% | 6 | 21% | 0 | 0% |
| Meredith | White | 33 | 25 | 4 | 30 | 18 | 72% | 8 | 32% | 14 | 56% | 12 | 48% | 1 | 4% |
| Molly | White | 70 | 49 | 0 | 52 | 26 | 53% | 12 | 24% | 28 | 57% | 10 | 20% | 1 | 2% |
| Neil | White | 30 | 20 | 5 | 21 | 18 | 90% | 7 | 35% | 6 | 30% | 3 | 15% | 2 | 10% |
| Nia | Black | 11 | 5 | 2 | 5 | 4 | 80% | 0 | 0% | 4 | 80% | 0 | 0% | 1 | 20% |
| Precious | Black | 12 | 7 | 2 | 8 | 2 | 29% | 3 | 43% | 2 | 29% | 2 | 29% | 1 | 14% |
| Rasheed | Black | 17 | 13 | 6 | 17 | 6 | 46% | 6 | 46% | 6 | 46% | 9 | 69% | 2 | 15% |
| Shanice | Black | 26 | 19 | 14 | 20 | 13 | 68% | 6 | 32% | 13 | 68% | 4 | 21% | 1 | 5% |
| Tamika | Black | 29 | 22 | 18 | 25 | 18 | 82% | 15 | 68% | 14 | 64% | 9 | 41% | 0 | 0% |
| Tanner | White | 30 | 18 | 2 | 20 | 15 | 83% | 1 | 6% | 9 | 50% | 3 | 17% | 0 | 0% |
| Terrell | Black | 28 | 24 | 18 | 24 | 20 | 83% | 11 | 46% | 14 | 58% | 11 | 46% | 1 | 4% |
| Tremayne | Black | 27 | 22 | 14 | 26 | 5 | 23% | 13 | 59% | 20 | 91% | 10 | 45% | 0 | 0% |
| Trevon | Black | 26 | 23 | 8 | 23 | 9 | 39% | 3 | 13% | 15 | 65% | 14 | 61% | 0 | 0% |
| Tyrone | Black | 36 | 36 | 33 | 36 | 30 | 83% | 25 | 69% | 19 | 53% | 11 | 31% | 1 | 3% |
| Wyatt | White | 30 | 22 | 4 | 23 | 11 | 50% | 7 | 32% | 11 | 50% | 6 | 27% | 0 | 0% |
| | Totals | | 1598 | 544 | **1705** | 1126 | | 550 | | 770 | | 535 | | 56 | |

**Figure 16. Counts of ads for public records by first name.**

| instantcheckmate | | peoplesmart | |
|---|---|---|---|
| C 382 | **Located: _fullname_** <br> Information found on _fullname fullname_ found in database. | A 7 | **We found: _fullname_** <br> 1) Get _firstname_'s Background Report 2) Contact info & More -try Free! |
| AC 2 | **Located: The Person** <br> Information found on them Person found in database. | T 87 | **We found: _fullname_** <br> 1) Get Aisha's Background Report 2) Current Contact Info - Try Free! |
| G* 96 | **We found _fullname_** <br> Search Arrests, Address, Phone, etc. Search records for _fullname_. | D 105 | **We found: _fullname_** <br> 1) Contact _fullname_ -Free Info! 2) Current Address, Phone & More. |
| S* 4 | **We found Them** <br> Search Arrests, Address, Phone, etc. Search records for _fullname_. | Q 348 | **We found: _fullname_** <br> 1) Contact _fullname_ -Free Info! 2) Current Phone, Address & More. |
| I 40 | **Background of _fullname_** <br> Search Instant Checkmate for the Records of _fullname_ | AG 1 | **We found _firstname_** <br> Get _firstname_ in CA's Email, Address, Phone, Public Records & More Easy! |
| U 9 | **Background of Anyone** <br> Search Instant Checkmate for the Records of _fullname_ | AH 1 | **We found _firstname_ In _lastname_** <br> 1)Get _firstname_'s Info – Try Now! 2)Current Phone, Address & More. |
| J 17 | **_fullname_'s Records** <br> 1) Enter Name and State. 2) Access Full Background Checks Instantly. | R 1 | **Looking For _fullname_?** <br> Get _fullname_'s Phone, Email Address, Public Records & More Now! |
| X 3 | **Anyone's Records** <br> 1) Enter Name and State. 2) Access Full Background Checks Instantly. | | |
| K* 195 | **_fullname_: Truth** <br> Arrests and Much More. Everything About _fullname_ | | **publicrecords** |
| O* 67 | **_fullname_ Truth** <br> Looking for _fullname_? Check _fullname_'s Arrests | B 570 | **_fullname_** <br> Public Records Found For: _fullname_. View now. |
| L* 176 | **_fullname_, Arrested?** <br> 1) Enter Name and State. 2) Access Full Background Checks Instantly. | P 128 | **_fullname_** <br> Public Records Found For: _fullname_. Search now. |
| V* 2 | **Uh Oh, Arrested?** <br> 1) Enter Name and State. 2) Access Full Background Checks Instantly. | F 13 | **Records: _fullname_** <br> Database of all _lastname_'s in the Country. Search now. |
| AD* 1 | **Found: _fullname_** <br> We have the story on _fullname fullname_'s arrests, relatives,etc. | Z 2 | **_Fullname_ Info** <br> View Contact Information For Free Quick & Easy Search Results! |
| AF* 3 | **_Fullname_ - Found** <br> Learn the truth about _fullname_ Check _fullname_'s arrests & more. | H 56 | **_fullname_** <br> We have Public Records For: _fullname_. Search Now. |
| AE 4 | **Research _fullname_** <br> We have details on _fullname_. _fullname_'s full background & info. | | |
| M* 55 | **_fullname_ Located** <br> Background Check, Arrest Records, Phone, & Address. Instant, Accurate | | **peoplefinders** |
| N 62 | **Looking for _fullname_?** <br> Comprehensive Background Report and More on _fullname_ | E 523 | **We found _fullname_** <br> Current Address, Phone and Age. Find _fullname_, Anywhere. |
| AI 8 | **Looking for People in the US?** <br> Comprehensive Background Report and More on _fullname_ | Y 8 | **We found _fullname_** <br> 1)Get Phone/ Address/ Age Instantly! 2) Find Anyone, Anywhere for Free. |
| | | AA 2 | **Find _fullname_** <br> Get current and past addresses and phone numbers. Instant results! |
| | | AB 1 | **We Found Them for Free** <br> Current  Address, Phone and Age. Find _fullname_ Anywhere. |

**Figure 17. Templates for ads for public records on Reuters, replace _fullname_ with person's first and last name.  Letter identifies text. Number is number of occurrences of text. *arrest ad.**

| FIRST NAME | | | INSTANT CHECKMATE | | | | | | | | | | | | | | | | | | PEOPLESMART | | | | | | | | PUBLICRECORDS | | | | | | PEOPLEFINDERS | | | |
| Name | Race | Full Names | C | AC | G | S | I | U | J | X | K | O | L | V | AD | AF | AE | M | N | AI | Totals | A | T | D | Q | AG | AH | R | Totals | B | P | F | Z | H | Totals | E | Y | AA | AB |
|---|---|---|---|---|---|---|---|---|---|---|---|---|---|---|---|---|---|---|---|---|---|---|---|---|---|---|---|---|---|---|---|---|---|---|---|---|---|---|---|
| Aaliyah | Black | 19 | 1 | 0 | 1 | 0 | 0 | 0 | 0 | 0 | 1 | 0 | 1 | 0 | 0 | 0 | 0 | 0 | 0 | 0 | 4 | 0 | 1 | 0 | 1 | 0 | 0 | 0 | 2 | 6 | 3 | 1 | 0 | 3 | 13 | 0 | 0 | 0 | 0 |
| Aisha | Black | 54 | 21 | 0 | 6 | 0 | 1 | 0 | 1 | 0 | 2 | 0 | 1 | 0 | 0 | 0 | 0 | 2 | 3 | 0 | 37 | 3 | 6 | 11 | 4 | 0 | 0 | 0 | 24 | 21 | 2 | 1 | 0 | 1 | 25 | 24 | 0 | 2 | 0 |
| Allison | White | 28 | 5 | 0 | 0 | 0 | 0 | 0 | 0 | 0 | 5 | 1 | 0 | 0 | 0 | 0 | 0 | 0 | 1 | 0 | 12 | 0 | 1 | 0 | 1 | 0 | 0 | 0 | 2 | 3 | 1 | 0 | 0 | 2 | 6 | 7 | 0 | 0 | 0 |
| Amy | White | 67 | 4 | 0 | 3 | 0 | 2 | 0 | 2 | 0 | 3 | 3 | 3 | 0 | 1 | 2 | 1 | 1 | 2 | 0 | 27 | 0 | 0 | 4 | 14 | 0 | 0 | 0 | 18 | 13 | 4 | 0 | 0 | 2 | 19 | 10 | 0 | 0 | 0 |
| Anne | White | 35 | 4 | 0 | 1 | 0 | 0 | 0 | 1 | 0 | 5 | 1 | 2 | 0 | 0 | 0 | 0 | 2 | 0 | 0 | 16 | 0 | 0 | 0 | 0 | 0 | 0 | 0 | 0 | 5 | 0 | 0 | 0 | 1 | 6 | 5 | 0 | 0 | 0 |
| Brad | White | 37 | 8 | 0 | 3 | 0 | 0 | 0 | 1 | 0 | 7 | 3 | 7 | 0 | 0 | 0 | 0 | 1 | 0 | 0 | 30 | 2 | 2 | 9 | 0 | 0 | 0 | 0 | 13 | 8 | 1 | 0 | 0 | 0 | 9 | 4 | 0 | 0 | 0 |
| Brendan | White | 40 | 13 | 0 | 2 | 0 | 0 | 1 | 0 | 0 | 8 | 1 | 2 | 0 | 0 | 0 | 0 | 1 | 5 | 1 | 34 | 0 | 7 | 3 | 0 | 0 | 0 | 0 | 10 | 13 | 2 | 0 | 0 | 0 | 15 | 6 | 0 | 0 | 0 |
| Brett | White | 28 | 12 | 0 | 1 | 0 | 0 | 0 | 0 | 1 | 1 | 2 | 2 | 0 | 0 | 0 | 0 | 1 | 1 | 0 | 21 | 0 | 1 | 5 | 0 | 0 | 0 | 0 | 6 | 8 | 1 | 0 | 0 | 0 | 9 | 4 | 0 | 0 | 0 |
| Carrie | White | 33 | 9 | 0 | 3 | 0 | 0 | 1 | 0 | 0 | 2 | 0 | 0 | 0 | 0 | 0 | 0 | 1 | 1 | 0 | 17 | 0 | 0 | 6 | 0 | 0 | 0 | 0 | 6 | 5 | 2 | 0 | 0 | 0 | 7 | 1 | 0 | 0 | 0 |
| Claire | White | 56 | 11 | 0 | 2 | 0 | 2 | 0 | 0 | 0 | 10 | 2 | 1 | 0 | 0 | 1 | 0 | 2 | 0 | 0 | 31 | 0 | 0 | 9 | 5 | 0 | 0 | 0 | 14 | 7 | 0 | 0 | 0 | 0 | 7 | 13 | 0 | 0 | 0 |
| Cody | White | 30 | 2 | 0 | 2 | 0 | 0 | 0 | 0 | 0 | 3 | 0 | 2 | 0 | 0 | 0 | 0 | 0 | 0 | 0 | 9 | 0 | 0 | 2 | 6 | 0 | 0 | 0 | 8 | 9 | 3 | 1 | 0 | 0 | 13 | 1 | 0 | 0 | 0 |
| Connor | White | 30 | 1 | 0 | 0 | 0 | 0 | 0 | 0 | 0 | 0 | 0 | 1 | 0 | 0 | 0 | 0 | 0 | 1 | 0 | 4 | 0 | 0 | 1 | 1 | 0 | 0 | 0 | 2 | 11 | 4 | 1 | 0 | 0 | 16 | 1 | 0 | 0 | 0 |
| Darnell | Black | 26 | 1 | 0 | 1 | 0 | 0 | 0 | 1 | 0 | 5 | 1 | 9 | 0 | 0 | 0 | 0 | 0 | 1 | 0 | 19 | 0 | 3 | 3 | 0 | 0 | 0 | 0 | 6 | 9 | 2 | 0 | 0 | 1 | 12 | 0 | 0 | 0 | 0 |
| DeAndre | Black | 29 | 4 | 0 | 0 | 0 | 0 | 0 | 0 | 0 | 4 | 1 | 3 | 0 | 0 | 0 | 0 | 2 | 1 | 0 | 15 | 0 | 1 | 3 | 2 | 0 | 0 | 0 | 6 | 12 | 1 | 0 | 0 | 0 | 13 | 8 | 0 | 0 | 0 |
| Deja | Black | 24 | 2 | 0 | 2 | 0 | 0 | 0 | 0 | 0 | 1 | 0 | 4 | 0 | 0 | 0 | 0 | 0 | 2 | 0 | 11 | 0 | 0 | 0 | 0 | 0 | 0 | 0 | 0 | 5 | 3 | 1 | 0 | 2 | 11 | 7 | 0 | 0 | 0 |
| DeShawn | Black | 27 | 1 | 0 | 8 | 1 | 1 | 0 | 0 | 0 | 2 | 1 | 5 | 0 | 0 | 0 | 0 | 0 | 1 | 0 | 21 | 0 | 1 | 1 | 5 | 0 | 0 | 0 | 7 | 14 | 1 | 0 | 0 | 1 | 16 | 1 | 0 | 0 | 0 |
| Diamond | Black | | | | | | | | | | | | | | | | | | | | | | | | | | | | | | | | | | | | | | |
| Dustin | White | 66 | 5 | 0 | 4 | 0 | 1 | 0 | 0 | 0 | 26 | 1 | 6 | 0 | 0 | 0 | 0 | 1 | 3 | 0 | 47 | 0 | 4 | 15 | 1 | 0 | 0 | 0 | 20 | 22 | 5 | 0 | 1 | 0 | 28 | 6 | 0 | 0 | 0 |
| Ebony | Black | 59 | 6 | 0 | 6 | 0 | 2 | 0 | 1 | 0 | 8 | 5 | 5 | 0 | 0 | 0 | 0 | 1 | 5 | 0 | 39 | 1 | 2 | 0 | 8 | 0 | 0 | 0 | 11 | 24 | 2 | 0 | 0 | 2 | 28 | 14 | 1 | 0 | 0 |
| Emily | White | 30 | 7 | 0 | 1 | 0 | 0 | 1 | 2 | 0 | 2 | 3 | 1 | 0 | 0 | 0 | 0 | 1 | 1 | 0 | 19 | 0 | 1 | 0 | 1 | 0 | 0 | 0 | 2 | 2 | 1 | 1 | 0 | 0 | 4 | 1 | 0 | 0 | 0 |
| Emma | White | 60 | 12 | 0 | 1 | 0 | 0 | 0 | 1 | 0 | 3 | 0 | 1 | 0 | 0 | 0 | 0 | 0 | 2 | 0 | 20 | 0 | 3 | 0 | 14 | 0 | 0 | 0 | 17 | 16 | 5 | 0 | 0 | 1 | 22 | 8 | 0 | 0 | 0 |
| Geoffrey | White | 34 | 17 | 0 | 2 | 0 | 0 | 0 | 0 | 0 | 1 | 0 | 1 | 0 | 0 | 0 | 0 | 3 | 0 | 0 | 24 | 0 | 2 | 1 | 6 | 0 | 0 | 0 | 9 | 7 | 4 | 0 | 0 | 2 | 13 | 5 | 0 | 0 | 0 |
| Greg | White | 40 | 18 | 0 | 1 | 0 | 1 | 0 | 0 | 0 | 4 | 0 | 6 | 0 | 0 | 0 | 0 | 2 | 3 | 0 | 35 | 0 | 1 | 0 | 5 | 0 | 0 | 0 | 6 | 7 | 3 | 0 | 0 | 0 | 10 | 12 | 2 | 0 | 0 |
| Hakim | Black | 17 | 1 | 0 | 1 | 0 | 0 | 0 | 0 | 0 | 3 | 0 | 0 | 0 | 0 | 0 | 0 | 0 | 0 | 0 | 5 | 0 | 0 | 2 | 0 | 0 | 0 | 0 | 2 | 2 | 1 | 0 | 0 | 1 | 4 | 5 | 0 | 0 | 0 |
| Imani | Black | 25 | 3 | 0 | 1 | 0 | 1 | 1 | 0 | 0 | 1 | 0 | 0 | 0 | 0 | 0 | 0 | 0 | 1 | 0 | 8 | 0 | 1 | 1 | 2 | 0 | 0 | 0 | 4 | 3 | 3 | 0 | 0 | 5 | 11 | 7 | 0 | 0 | 0 |
| Jack | White | 58 | 14 | 0 | 0 | 0 | 1 | 0 | 0 | 0 | 6 | 1 | 2 | 0 | 0 | 0 | 0 | 0 | 1 | 0 | 25 | 0 | 2 | 7 | 0 | 0 | 0 | 0 | 9 | 12 | 1 | 0 | 0 | 0 | 13 | 9 | 1 | 0 | 0 |
| Jake | White | 59 | 9 | 0 | 1 | 0 | 1 | 0 | 0 | 0 | 3 | 0 | 1 | 0 | 0 | 0 | 1 | 3 | 1 | 0 | 20 | 0 | 0 | 4 | 4 | 0 | 0 | 0 | 8 | 5 | 3 | 0 | 0 | 0 | 8 | 8 | 0 | 0 | 0 |
| Jamal | Black | 29 | 5 | 0 | 4 | 0 | 0 | 0 | 0 | 0 | 1 | 0 | 0 | 0 | 0 | 0 | 0 | 0 | 1 | 0 | 11 | 0 | 2 | 1 | 6 | 0 | 0 | 0 | 9 | 7 | 0 | 0 | 0 | 1 | 8 | 16 | 0 | 0 | 0 |
| Jay | White | 29 | 8 | 0 | 1 | 2 | 0 | 0 | 0 | 0 | 4 | 0 | 2 | 0 | 0 | 0 | 0 | 0 | 1 | 0 | 18 | 0 | 1 | 0 | 1 | 0 | 0 | 0 | 2 | 3 | 0 | 0 | 0 | 0 | 3 | 4 | 1 | 0 | 0 |
| Jermaine | Black | 28 | 2 | 0 | 0 | 0 | 0 | 0 | 0 | 0 | 3 | 0 | 9 | 1 | 0 | 0 | 0 | 0 | 1 | 0 | 16 | 0 | 3 | 3 | 8 | 0 | 0 | 0 | 14 | 13 | 0 | 0 | 0 | 0 | 13 | 6 | 0 | 0 | 1 |
| Jill | White | 34 | 18 | 0 | 3 | 0 | 2 | 0 | 0 | 0 | 3 | 0 | 0 | 0 | 0 | 0 | 0 | 0 | 0 | 0 | 26 | 0 | 1 | 3 | 3 | 0 | 0 | 0 | 7 | 7 | 3 | 1 | 0 | 0 | 11 | 8 | 0 | 0 | 0 |
| Kareem | Black | 33 | 9 | 0 | 0 | 0 | 0 | 0 | 1 | 0 | 3 | 0 | 4 | 0 | 0 | 0 | 0 | 0 | 4 | 0 | 21 | 0 | 5 | 0 | 6 | 0 | 0 | 0 | 11 | 12 | 1 | 0 | 0 | 0 | 13 | 6 | 0 | 0 | 0 |
| Katelyn | White | 80 | 12 | 0 | 2 | 0 | 2 | 0 | 0 | 0 | 3 | 2 | 2 | 0 | 0 | 0 | 0 | 1 | 1 | 0 | 26 | 0 | 2 | 6 | 0 | 0 | 0 | 0 | 8 | 19 | 8 | 0 | 0 | 7 | 34 | 54 | 0 | 0 | 0 |
| Katie | White | 50 | 8 | 0 | 0 | 0 | 1 | 0 | 0 | 0 | 2 | 2 | 0 | 0 | 0 | 0 | 0 | 1 | 0 | 0 | 14 | 0 | 0 | 1 | 12 | 0 | 0 | 0 | 13 | 7 | 3 | 0 | 0 | 0 | 10 | 7 | 0 | 0 | 0 |
| Keisha | Black | 40 | 12 | 0 | 2 | 0 | 2 | 0 | 0 | 0 | 3 | 1 | 3 | 0 | 0 | 0 | 0 | 0 | 1 | 0 | 24 | 3 | 3 | 3 | 6 | 0 | 0 | 1 | 16 | 8 | 0 | 0 | 0 | 0 | 8 | 13 | 0 | 0 | 0 |
| Kenya | Black | 4 | 1 | 0 | 1 | 0 | 0 | 0 | 1 | 0 | 1 | 0 | 0 | 0 | 0 | 0 | 0 | 0 | 0 | 0 | 4 | 0 | 1 | 0 | 0 | 0 | 0 | 0 | 1 | 1 | 0 | 0 | 0 | 0 | 1 | 3 | 0 | 0 | 0 |
| Kristen | White | 34 | 2 | 0 | 0 | 0 | 0 | 0 | 0 | 0 | 0 | 0 | 1 | 0 | 0 | 0 | 0 | 0 | 0 | 0 | 3 | 0 | 3 | 0 | 7 | 0 | 0 | 0 | 10 | 6 | 3 | 0 | 0 | 2 | 11 | 7 | 0 | 0 | 0 |
| Lakisha | Black | 28 | 6 | 0 | 1 | 0 | 1 | 0 | 0 | 0 | 0 | 0 | 1 | 0 | 0 | 0 | 0 | 0 | 0 | 0 | 10 | 0 | 2 | 1 | 12 | 0 | 0 | 0 | 15 | 5 | 0 | 0 | 0 | 0 | 5 | 17 | 0 | 0 | 0 |
| Latanya | Black | 28 | 11 | 0 | 0 | 0 | 0 | 0 | 0 | 0 | 1 | 0 | 7 | 0 | 0 | 0 | 0 | 0 | 0 | 0 | 19 | 0 | 6 | 4 | 9 | 0 | 0 | 0 | 19 | 11 | 2 | 0 | 0 | 0 | 13 | 4 | 0 | 0 | 0 |
| Latisha | Black | 28 | 5 | 0 | 3 | 0 | 1 | 0 | 0 | 0 | 1 | 1 | 5 | 0 | 0 | 0 | 0 | 0 | 1 | 0 | 17 | 0 | 0 | 5 | 0 | 0 | 0 | 0 | 5 | 11 | 1 | 0 | 0 | 0 | 12 | 20 | 0 | 0 | 0 |
| Latonya | Black | 36 | 6 | 0 | 2 | 1 | 1 | 0 | 1 | 0 | 1 | 4 | 8 | 0 | 0 | 0 | 0 | 1 | 1 | 0 | 26 | 0 | 0 | 17 | 0 | 0 | 0 | 0 | 17 | 15 | 2 | 0 | 0 | 0 | 17 | 13 | 1 | 0 | 0 |
| Latoya | Black | 27 | 6 | 0 | 3 | 0 | 1 | 0 | 0 | 0 | 5 | 2 | 6 | 0 | 0 | 0 | 0 | 1 | 1 | 0 | 25 | 0 | 0 | 9 | 0 | 0 | 0 | 0 | 9 | 10 | 4 | 0 | 0 | 1 | 15 | 6 | 0 | 0 | 0 |
| Laurie | White | 28 | 2 | 0 | 2 | 0 | 0 | 1 | 0 | 0 | 5 | 0 | 3 | 0 | 0 | 0 | 0 | 1 | 0 | 0 | 14 | 0 | 2 | 0 | 3 | 0 | 0 | 0 | 5 | 6 | 2 | 0 | 0 | 0 | 8 | 6 | 0 | 0 | 0 |
| Leroy | Black | 25 | 6 | 1 | 4 | 1 | 1 | 0 | 0 | 0 | 3 | 0 | 2 | 0 | 0 | 0 | 0 | 2 | 2 | 0 | 22 | 0 | 1 | 0 | 12 | 0 | 0 | 0 | 13 | 7 | 3 | 0 | 0 | 1 | 11 | 4 | 1 | 0 | 0 |
| Luke | White | 60 | 7 | 0 | 2 | 0 | 4 | 1 | 0 | 0 | 1 | 3 | 3 | 0 | 0 | 0 | 0 | 1 | 1 | 0 | 23 | 0 | 3 | 14 | 0 | 1 | 0 | 0 | 18 | 12 | 4 | 0 | 0 | 2 | 18 | 4 | 0 | 0 | 0 |
| Madeline | White | 66 | 2 | 0 | 0 | 0 | 0 | 1 | 0 | 1 | 1 | 1 | 0 | 0 | 0 | 0 | 0 | 1 | 3 | 0 | 11 | 0 | 0 | 7 | 10 | 0 | 0 | 0 | 17 | 17 | 5 | 0 | 0 | 2 | 24 | 36 | 0 | 0 | 0 |
| Malik | Black | 18 | 2 | 0 | 0 | 0 | 1 | 0 | 0 | 0 | 1 | 1 | 2 | 0 | 0 | 0 | 0 | 0 | 1 | 0 | 8 | 0 | 0 | 2 | 0 | 0 | 0 | 0 | 2 | 3 | 0 | 0 | 0 | 1 | 4 | 7 | 0 | 0 | 0 |
| Marquis | Black | 19 | 4 | 0 | 1 | 0 | 0 | 0 | 0 | 0 | 0 | 1 | 8 | 0 | 0 | 0 | 0 | 0 | 0 | 0 | 14 | 0 | 1 | 1 | 3 | 0 | 0 | 0 | 5 | 8 | 1 | 0 | 0 | 0 | 9 | 0 | 0 | 0 | 0 |
| Matthew | White | 44 | 8 | 0 | 2 | 0 | 0 | 0 | 2 | 0 | 7 | 0 | 4 | 0 | 0 | 0 | 0 | 1 | 0 | 0 | 24 | 0 | 3 | 1 | 1 | 0 | 0 | 0 | 5 | 7 | 2 | 1 | 0 | 0 | 10 | 5 | 1 | 0 | 0 |
| Meredith | White | 33 | 9 | 0 | 0 | 0 | 0 | 2 | 0 | 0 | 2 | 2 | 1 | 0 | 0 | 0 | 0 | 1 | 0 | 1 | 18 | 0 | 3 | 1 | 4 | 0 | 0 | 0 | 8 | 12 | 2 | 0 | 0 | 0 | 14 | 12 | 0 | 0 | 0 |
| Molly | White | 70 | 6 | 0 | 1 | 0 | 1 | 0 | 1 | 0 | 9 | 6 | 0 | 0 | 0 | 0 | 0 | 1 | 1 | 0 | 26 | 0 | 2 | 10 | 0 | 0 | 0 | 0 | 12 | 18 | 8 | 0 | 0 | 2 | 28 | 10 | 0 | 0 | 0 |
| Neil | White | 30 | 6 | 0 | 4 | 0 | 1 | 0 | 0 | 0 | 2 | 2 | 2 | 0 | 0 | 0 | 0 | 0 | 1 | 0 | 18 | 0 | 2 | 1 | 4 | 0 | 0 | 0 | 7 | 5 | 1 | 0 | 0 | 0 | 6 | 9 | 0 | 0 | 0 |
| Nia | Black | 11 | 1 | 1 | 0 | 0 | 0 | 0 | 0 | 0 | 2 | 0 | 0 | 0 | 0 | 0 | 0 | 0 | 0 | 0 | 4 | 0 | 0 | 0 | 0 | 0 | 0 | 0 | 0 | 2 | 0 | 1 | 0 | 1 | 4 | 4 | 0 | 0 | 0 |
| Precious | Black | 12 | 0 | 0 | 0 | 0 | 0 | 0 | 0 | 0 | 0 | 1 | 0 | 0 | 0 | 0 | 0 | 0 | 0 | 0 | 2 | 0 | 2 | 0 | 0 | 0 | 0 | 0 | 2 | 3 | 0 | 0 | 0 | 0 | 3 | 2 | 0 | 0 | 0 |
| Rasheed | Black | 17 | 1 | 0 | 0 | 0 | 0 | 0 | 0 | 0 | 1 | 0 | 0 | 0 | 0 | 0 | 0 | 0 | 1 | 0 | 6 | 0 | 3 | 3 | 0 | 0 | 0 | 0 | 6 | 6 | 0 | 0 | 0 | 0 | 6 | 9 | 0 | 0 | 0 |
| Shanice | Black | 26 | 6 | 0 | 0 | 0 | 0 | 0 | 0 | 0 | 3 | 0 | 4 | 0 | 0 | 0 | 0 | 0 | 0 | 0 | 13 | 0 | 3 | 0 | 0 | 0 | 0 | 0 | 3 | 9 | 1 | 0 | 0 | 3 | 13 | 8 | 0 | 0 | 0 |
| Tamika | Black | 29 | 5 | 0 | 1 | 0 | 2 | 0 | 0 | 0 | 0 | 0 | 2 | 0 | 0 | 0 | 0 | 0 | 5 | 0 | 15 | 0 | 2 | 7 | 6 | 0 | 0 | 0 | 15 | 9 | 3 | 1 | 0 | 0 | 13 | 8 | 0 | 0 | 0 |
| Tanner | White | 30 | 6 | 0 | 0 | 0 | 0 | 0 | 0 | 0 | 1 | 1 | 2 | 0 | 0 | 0 | 0 | 0 | 1 | 0 | 11 | 0 | 0 | 1 | 0 | 0 | 0 | 0 | 1 | 6 | 3 | 0 | 0 | 0 | 9 | 4 | 0 | 0 | 0 |
| Terrell | Black | 28 | 1 | 0 | 1 | 0 | 3 | 0 | 0 | 0 | 3 | 0 | 9 | 0 | 0 | 0 | 0 | 0 | 2 | 0 | 19 | 0 | 2 | 7 | 0 | 0 | 0 | 0 | 9 | 11 | 2 | 0 | 0 | 1 | 14 | 10 | 0 | 0 | 0 |
| Tremayne | Black | 27 | 0 | 0 | 0 | 0 | 0 | 0 | 0 | 0 | 1 | 2 | 0 | 0 | 0 | 0 | 0 | 0 | 1 | 1 | 5 | 0 | 4 | 3 | 6 | 0 | 0 | 0 | 13 | 15 | 2 | 2 | 0 | 2 | 21 | 11 | 0 | 0 | 0 |
| Trevon | Black | 26 | 2 | 0 | 2 | 0 | 0 | 0 | 0 | 0 | 0 | 2 | 0 | 0 | 0 | 0 | 0 | 2 | 1 | 0 | 10 | 0 | 0 | 1 | 1 | 2 | 0 | 0 | 4 | 8 | 3 | 1 | 1 | 1 | 14 | 13 | 0 | 0 | 0 |
| Tyrone | Black | 36 | 3 | 0 | 1 | 0 | 0 | 1 | 0 | 0 | 8 | 1 | 13 | 0 | 0 | 0 | 0 | 0 | 1 | 2 | 0 | 0 | 3 | 4 | 17 | 0 | 0 | 0 | 24 | 20 | 11 | 0 | 0 | 0 | | 11 | 0 | 0 | 0 |
| Wyatt | White | 30 | 3 | 0 | 1 | 0 | 1 | 0 | 0 | 0 | 2 | 0 | 1 | 0 | 0 | 0 | 0 | 0 | 0 | 0 | 9 | 0 | 1 | 0 | 2 | 0 | 0 | 0 | 3 | | | | | | | | | | |
| **Totals** | | | 382 | 2 | 96 | 4 | 40 | 9 | 17 | 3 | 195 | 67 | 176 | 2 | 1 | 3 | 4 | 55 | 62 | 8 | 1126 | 7 | 87 | 105 | 348 | 1 | 1 | 1 | 550 | 570 | 128 | 13 | 2 | 56 | 769 | 523 | 8 | 2 | 1 |
| | | | | | | | | | | | | | | | | | | | | | 1126 | | | | | | | | 550 | | | | | | 769 | | | | 534 |

**Figure 18. Distribution of ad templates in Figure 17 by first name as they appeared on Reuters.com.**

**INSTANT CHECKMATE ADS ON REUTERS**

| Name | Race | Total Ads | Arrest Ads | | Neutral Ads | |
|------|------|-----------|------|------|------|------|
| Aaliyah | Black | 4 | 3 | 75% | 1 | 25% |
| Aisha | Black | 37 | 11 | 30% | 26 | 70% |
| Allison | White | 12 | 6 | 50% | 6 | 50% |
| Amy | White | 27 | 16 | 59% | 11 | 41% |
| Anne | White | 16 | 11 | 69% | 5 | 31% |
| Brad | White | 30 | 21 | 70% | 9 | 30% |
| Brendan | White | 34 | 14 | 41% | 20 | 59% |
| Brett | White | 21 | 7 | 33% | 14 | 67% |
| Carrie | White | 17 | 6 | 35% | 11 | 65% |
| Claire | White | 31 | 18 | 58% | 13 | 42% |
| Cody | White | 9 | 7 | 78% | 2 | 22% |
| Connor | White | 4 | 2 | 50% | 2 | 50% |
| Darnell | Black | 19 | 16 | 84% | 3 | 16% |
| DeAndre | Black | 15 | 10 | 67% | 5 | 33% |
| Deja | Black | 11 | 7 | 64% | 4 | 36% |
| DeShawn | Black | 21 | 18 | 86% | 3 | 14% |
| Diamond | Black |  |  |  |  |  |
| Dustin | White | 47 | 38 | 81% | 9 | 19% |
| Ebony | Black | 39 | 25 | 64% | 14 | 36% |
| Emily | White | 19 | 8 | 42% | 11 | 58% |
| Emma | White | 20 | 5 | 25% | 15 | 75% |
| Geoffrey | White | 24 | 7 | 29% | 17 | 71% |
| Greg | White | 35 | 13 | 37% | 22 | 63% |
| Hakim | Black | 5 | 4 | 80% | 1 | 20% |
| Imani | Black | 8 | 2 | 25% | 6 | 75% |
| Jack | White | 25 | 9 | 36% | 16 | 64% |
| Jake | White | 20 | 8 | 40% | 12 | 60% |
| Jamal | Black | 11 | 5 | 45% | 6 | 55% |
| Jay | White | 18 | 7 | 39% | 11 | 61% |
| Jermaine | Black | 16 | 13 | 81% | 3 | 19% |
| Jill | White | 26 | 6 | 23% | 20 | 77% |
| Kareem | Black | 21 | 7 | 33% | 14 | 67% |
| Katelyn | White | 26 | 10 | 38% | 16 | 62% |
| Katie | White | 14 | 5 | 36% | 9 | 64% |
| Keisha | Black | 24 | 9 | 38% | 15 | 63% |
| Kenya | Black | 4 | 2 | 50% | 2 | 50% |
| Kristen | White | 3 | 1 | 33% | 2 | 67% |
| Lakisha | Black | 15 | 8 | 53% | 7 | 47% |
| Latanya | Black | 19 | 8 | 42% | 11 | 58% |
| Latisha | Black | 17 | 11 | 65% | 6 | 35% |
| Latonya | Black | 26 | 17 | 65% | 9 | 35% |
| Latoya | Black | 25 | 17 | 68% | 8 | 32% |
| Laurie | White | 14 | 11 | 79% | 3 | 21% |
| Leroy | Black | 22 | 12 | 55% | 10 | 45% |
| Luke | White | 23 | 10 | 43% | 13 | 57% |
| Madeline | White | 11 | 5 | 45% | 6 | 55% |
| Malik | Black | 8 | 4 | 50% | 4 | 50% |
| Marquis | Black | 14 | 10 | 71% | 4 | 29% |
| Matthew | White | 24 | 14 | 58% | 10 | 42% |
| Meredith | White | 18 | 6 | 33% | 12 | 67% |
| Molly | White | 26 | 17 | 65% | 9 | 35% |
| Neil | White | 18 | 11 | 61% | 7 | 39% |
| Nia | Black | 4 | 2 | 50% | 2 | 50% |
| Precious | Black | 2 | 1 | 50% | 1 | 50% |
| Rasheed | Black | 6 | 4 | 67% | 2 | 33% |
| Shanice | Black | 13 | 7 | 54% | 6 | 46% |
| Tamika | Black | 18 | 10 | 56% | 8 | 44% |
| Tanner | White | 16 | 5 | 31% | 11 | 69% |
| Terrell | Black | 19 | 15 | 79% | 4 | 21% |
| Tremayne | Black | 5 | 2 | 40% | 3 | 60% |
| Trevon | Black | 10 | 7 | 70% | 3 | 30% |
| Tyrone | Black | 30 | 24 | 80% | 6 | 20% |
| Wyatt | White | 10 | 4 | 40% | 6 | 60% |
| **Totals** | | **1126** | **599** | **53%** | **527** | **47%** |

| ARREST ADS SORT | | NEUTRAL ADS SORT | |
|------|------|------|------|
| 86% | DeShawn | 77% | Jill |
| 84% | Darnell | 75% | Emma |
| 81% | Jermaine | 75% | Imani |
| 81% | Dustin | 71% | Geoffrey |
| 80% | Hakim | 70% | Aisha |
| 80% | Tyrone | 69% | Tanner |
| 79% | Terrell | 67% | Brett |
| 79% | Laurie | 67% | Kareem |
| 78% | Cody | 67% | Kristen |
| 75% | Aaliyah | 67% | Meredith |
| 71% | Marquis | 65% | Carrie |
| 70% | Brad | 64% | Jack |
| 70% | Trevon | 63% | Greg |
| 69% | Anne | 63% | Keisha |
| 68% | Latoya | 62% | Katelyn |
| 67% | DeAndre | 61% | Jay |
| 67% | Rasheed | 60% | Jake |
| 65% | Latonya | 60% | Tremayne |
| 65% | Molly | 60% | Wyatt |
| 65% | Latisha | 59% | Brendan |
| 64% | Ebony | 58% | Emily |
| 64% | Deja | 58% | Latanya |
| 61% | Neil | 57% | Luke |
| 59% | Amy | 55% | Jamal |
| 58% | Matthew | 55% | Madeline |
| 58% | Claire | 50% | Allison |
| 56% | Tamika | 50% | Connor |
| 55% | Leroy | 50% | Kenya |
| 54% | Shanice | 50% | Malik |
| 53% | Lakisha | 50% | Nia |
| 50% | Allison | 50% | Precious |
| 50% | Connor | 47% | Lakisha |
| 50% | Kenya | 46% | Shanice |
| 50% | Malik | 45% | Leroy |
| 50% | Nia | 44% | Tamika |
| 50% | Precious | 42% | Claire |
| 45% | Jamal | 42% | Matthew |
| 45% | Madeline | 41% | Amy |
| 43% | Luke | 39% | Neil |
| 42% | Emily | 36% | Deja |
| 42% | Latanya | 36% | Ebony |
| 41% | Brendan | 35% | Latisha |
| 40% | Jake | 35% | Latonya |
| 40% | Tremayne | 35% | Molly |
| 40% | Wyatt | 33% | DeAndre |
| 39% | Jay | 33% | Rasheed |
| 38% | Katelyn | 32% | Latoya |
| 38% | Keisha | 31% | Anne |
| 37% | Greg | 30% | Brad |
| 36% | Jack | 30% | Trevon |
| 36% | Katie | 29% | Marquis |
| 35% | Carrie | 25% | Aaliyah |
| 33% | Brett | 22% | Cody |
| 33% | Kareem | 21% | Terrell |
| 33% | Kristen | 20% | Hakim |
| 33% | Meredith | 20% | Tyrone |
| 31% | Tanner | 19% | Dustin |
| 30% | Aisha | 19% | Jermaine |
| 29% | Geoffrey | 16% | Darnell |
| 25% | Emma | 14% | DeShawn |
| 25% | Imani |  | Diamond |
| 23% | Jill |  |  |
|  | Diamond |  |  |

**Figure 19. Distributions of Instant Checkmate ads having the word "arrest" or not ("neutral") appearing on Reuters.com.**

| Ad Templates on Reuters and Google | | Ad Templates on Google Only | |
|---|---|---|---|
| C 33 | **Located: *fullname*** <br> Information found on *fullname fullname* found in database. | AJ* 30 | ***fullname*'s Records** <br> Did you know *fullname*'s criminal history is searchable? |
| G* 24 | **We found *fullname*** <br> Search Arrests, Address, Phone, etc. Search records for *fullname*. | AP* 2 | ***fullname*'s Records Online?** <br> Did you know *fullname*'s criminal history is searchable? |
| I 2 | **Background of *fullname*** <br> Search Instant Checkmate for the Records of *fullname* | AM* 9 | **Anyone's Records Online?** <br> Did you know *fullname*'s criminal history is searchable? |
| U 1 | **Background of Anyone** <br> Search Instant Checkmate for the Records of *fullname* | AK* 2 | **Records For Anyone in US** <br> View Anyone's Criminal History. Check Criminal Records in Seconds! |
| J 6 | ***fullname*'s Records** <br> 1) Enter Name and State. 2) Access Full Background Checks Instantly. | AN* 9 | **Records For *fullname*** <br> View Anyone's Criminal History. Check Criminal Records in Seconds! |
| K* 52 | ***fullname*: Truth** <br> Arrests and Much More. Everything About *fullname* | AL* 26 | **Records For *fullname*?** <br> Find the Truth About *fullname* View Criminal Records in Seconds. |
| O* 7 | ***fullname* Truth** <br> Looking for *fullname*? Check *fullname*'s Arrests | AQ* 3 | **Records For People in the US?** <br> Find the Truth About *fullname* View Criminal Records in Seconds. |
| L* 200 | ***fullname*, Arrested?** <br> 1) Enter Name and State. 2) Access Full Background Checks Instantly. | AO* 6 | **Find *fullname*** <br> Criminal records, phone, address, & more on *fullname* |
| V* 10 | **Uh Oh, Arrested?** <br> 1) Enter Name and State. 2) Access Full Background Checks Instantly. | AS* 1 | **We Found *fullname* \| InstantCheckmate.com** <br> Search Arrests, Address, Phone, etc Search records for *fullname*. |
| M* 6 | ***fullname* Located** <br> Background Check, Arrest Records, Phone, & Address. Instant, Accurate | AT* 1 | ***fullname*'s Records \| InstantCheckmate.com** <br> Did you know *fullname*'s criminal history is searchable? |
| N 2 | **Looking for *fullname*?** <br> Comprehensive Background Report and More on *fullname* | | |

**Figure 20. Templates for ads for public records on Google.com, replace *fullname* with person's first and last name. Letter identifies text. Number is number of occurrences of text. Asterisk (*) denotes an ad suggestive of an arrest record.**

| | FIRST NAME | | INSTANT CHECKMATE ADS ON GOOGLE | | | | | | | | | | | | | | | | | | | | | | | | | | | |
|---|---|---|---|---|---|---|---|---|---|---|---|---|---|---|---|---|---|---|---|---|---|---|---|---|---|---|---|---|---|---|---|
| Name | Race | Full Names | C | AC | G | S | I | U | J | X | K | O | L | V | AD | AF | AE | M | N | AI | AJ | AP | AM | AK | AN | AL | AQ | AO | AS | AT | Totals |
| Aaliyah | Black | 19 | 0 | 0 | 0 | 0 | 0 | 0 | 0 | 0 | 0 | 0 | 0 | 0 | 0 | 0 | 0 | 0 | 0 | 0 | 0 | 0 | 0 | 0 | 0 | 0 | 0 | 0 | 0 | 0 | 0 | 0 |
| Aisha | Black | 54 | 4 | 0 | 0 | 0 | 0 | 0 | 0 | 0 | 3 | 1 | 3 | 1 | 0 | 0 | 0 | 4 | 0 | 0 | 1 | 1 | 1 | 0 | 0 | 4 | 0 | 0 | 0 | 0 | 23 |
| Allison | White | 28 | 0 | 0 | 0 | 0 | 0 | 0 | 0 | 0 | 0 | 0 | 0 | 0 | 0 | 0 | 0 | 0 | 0 | 0 | 0 | 0 | 0 | 0 | 0 | 0 | 0 | 0 | 0 | 0 | 0 |
| Amy | White | 67 | 0 | 0 | 0 | 0 | 0 | 0 | 0 | 0 | 0 | 0 | 0 | 0 | 0 | 0 | 0 | 0 | 0 | 0 | 0 | 0 | 0 | 0 | 0 | 0 | 0 | 0 | 0 | 0 | 0 |
| Anne | White | 35 | 0 | 0 | 0 | 0 | 0 | 0 | 0 | 0 | 0 | 0 | 0 | 0 | 0 | 0 | 0 | 0 | 0 | 0 | 0 | 0 | 0 | 0 | 0 | 0 | 0 | 0 | 0 | 0 | 0 |
| Brad | White | 37 | 0 | 0 | 1 | 0 | 0 | 0 | 0 | 0 | 0 | 0 | 1 | 5 | 0 | 0 | 0 | 0 | 0 | 0 | 0 | 0 | 0 | 0 | 0 | 0 | 0 | 1 | 0 | 0 | 8 |
| Brendan | White | 40 | 1 | 0 | 0 | 0 | 0 | 0 | 0 | 0 | 0 | 0 | 6 | 0 | 0 | 0 | 0 | 0 | 0 | 0 | 1 | 0 | 0 | 0 | 1 | 1 | 1 | 0 | 0 | 0 | 11 |
| Brett | White | 28 | 1 | 0 | 0 | 0 | 0 | 0 | 0 | 0 | 0 | 0 | 1 | 0 | 0 | 0 | 0 | 0 | 0 | 0 | 0 | 0 | 0 | 0 | 0 | 1 | 0 | 0 | 0 | 0 | 3 |
| Carrie | White | 33 | 1 | 0 | 0 | 0 | 0 | 0 | 0 | 0 | 0 | 0 | 1 | 0 | 0 | 0 | 0 | 0 | 0 | 0 | 0 | 0 | 0 | 0 | 0 | 0 | 0 | 0 | 0 | 0 | 2 |
| Claire | White | 56 | 0 | 0 | 0 | 0 | 0 | 0 | 0 | 0 | 0 | 0 | 0 | 0 | 0 | 0 | 0 | 0 | 0 | 0 | 0 | 0 | 0 | 0 | 0 | 0 | 0 | 0 | 0 | 0 | 0 |
| Cody | White | 30 | 0 | 0 | 0 | 0 | 0 | 0 | 0 | 0 | 0 | 0 | 0 | 0 | 0 | 0 | 0 | 0 | 0 | 0 | 0 | 0 | 0 | 0 | 0 | 0 | 0 | 0 | 0 | 0 | 0 |
| Connor | White | 30 | 0 | 0 | 0 | 0 | 0 | 0 | 0 | 0 | 0 | 0 | 0 | 0 | 0 | 0 | 0 | 0 | 0 | 0 | 0 | 0 | 0 | 0 | 0 | 0 | 0 | 0 | 0 | 0 | 0 |
| Darnell | Black | 26 | 0 | 0 | 0 | 0 | 0 | 1 | 0 | 0 | 0 | 0 | 13 | 0 | 0 | 0 | 0 | 0 | 0 | 0 | 2 | 0 | 1 | 0 | 0 | 1 | 0 | 0 | 0 | 0 | 18 |
| DeAndre | Black | 29 | 0 | 0 | 0 | 0 | 0 | 0 | 0 | 0 | 3 | 1 | 2 | 1 | 0 | 0 | 0 | 0 | 0 | 0 | 0 | 0 | 0 | 0 | 0 | 0 | 0 | 0 | 0 | 0 | 7 |
| Deja | Black | 24 | 0 | 0 | 2 | 0 | 0 | 0 | 0 | 0 | 1 | 0 | 4 | 0 | 0 | 0 | 0 | 0 | 1 | 0 | 0 | 0 | 0 | 0 | 0 | 2 | 0 | 0 | 0 | 0 | 10 |
| DeShawn | Black | 27 | 0 | 0 | 0 | 0 | 0 | 0 | 1 | 0 | 3 | 0 | 6 | 0 | 0 | 0 | 0 | 0 | 0 | 0 | 2 | 0 | 0 | 0 | 0 | 0 | 0 | 0 | 0 | 0 | 12 |
| Diamond | Black | | | | | | | | | | | | | | | | | | | | | | | | | | | | | | |
| Dustin | White | 66 | 0 | 0 | 0 | 0 | 0 | 0 | 0 | 0 | 0 | 0 | 1 | 0 | 0 | 0 | 0 | 0 | 0 | 0 | 0 | 0 | 0 | 0 | 0 | 0 | 0 | 0 | 0 | 0 | 1 |
| Ebony | Black | 59 | 1 | 0 | 0 | 0 | 0 | 1 | 0 | 0 | 3 | 0 | 17 | 0 | 0 | 0 | 0 | 0 | 0 | 0 | 3 | 0 | 1 | 0 | 0 | 0 | 0 | 0 | 0 | 0 | 26 |
| Emily | White | 30 | 0 | 0 | 0 | 0 | 0 | 0 | 0 | 0 | 0 | 0 | 0 | 0 | 0 | 0 | 0 | 0 | 0 | 0 | 0 | 0 | 0 | 0 | 0 | 0 | 0 | 0 | 0 | 0 | 0 |
| Emma | White | 60 | 0 | 0 | 0 | 0 | 0 | 0 | 0 | 0 | 0 | 0 | 0 | 0 | 0 | 0 | 0 | 0 | 0 | 0 | 0 | 0 | 0 | 0 | 0 | 0 | 0 | 0 | 0 | 0 | 0 |
| Geoffrey | White | 34 | 2 | 0 | 1 | 0 | 0 | 0 | 0 | 0 | 0 | 0 | 0 | 0 | 0 | 0 | 0 | 0 | 0 | 0 | 1 | 0 | 0 | 0 | 0 | 1 | 0 | 0 | 0 | 0 | 5 |
| Greg | White | 40 | 1 | 0 | 0 | 0 | 0 | 0 | 0 | 0 | 1 | 0 | 1 | 0 | 0 | 0 | 0 | 0 | 0 | 0 | 1 | 0 | 0 | 0 | 1 | 1 | 0 | 0 | 0 | 0 | 6 |
| Hakim | Black | 17 | 0 | 0 | 0 | 0 | 0 | 0 | 0 | 0 | 1 | 0 | 1 | 0 | 0 | 0 | 0 | 0 | 0 | 0 | 1 | 0 | 0 | 0 | 0 | 0 | 0 | 0 | 0 | 0 | 3 |
| Imani | Black | 25 | 0 | 0 | 0 | 0 | 0 | 0 | 0 | 0 | 1 | 0 | 1 | 0 | 0 | 0 | 0 | 0 | 0 | 0 | 1 | 0 | 0 | 0 | 0 | 0 | 0 | 0 | 0 | 0 | 3 |
| Jack | White | 58 | 0 | 0 | 0 | 0 | 0 | 0 | 0 | 0 | 0 | 0 | 0 | 0 | 0 | 0 | 0 | 0 | 0 | 0 | 0 | 0 | 0 | 0 | 0 | 0 | 0 | 0 | 0 | 0 | 0 |
| Jake | White | 59 | 0 | 0 | 0 | 0 | 0 | 0 | 0 | 0 | 0 | 0 | 0 | 0 | 0 | 0 | 0 | 0 | 0 | 0 | 0 | 0 | 0 | 0 | 0 | 0 | 0 | 0 | 0 | 0 | 0 |
| Jamal | Black | 29 | 1 | 0 | 1 | 0 | 0 | 0 | 0 | 0 | 0 | 0 | 0 | 0 | 0 | 0 | 0 | 0 | 0 | 0 | 0 | 0 | 0 | 2 | 1 | 0 | 1 | 0 | 0 | 0 | 6 |
| Jay | White | 29 | 0 | 0 | 1 | 0 | 0 | 0 | 0 | 0 | 1 | 0 | 1 | 0 | 0 | 0 | 0 | 0 | 0 | 0 | 0 | 0 | 0 | 0 | 1 | 0 | 0 | 0 | 0 | 0 | 4 |
| Jermaine | Black | 28 | 0 | 0 | 1 | 0 | 0 | 0 | 1 | 0 | 1 | 0 | 10 | 1 | 0 | 0 | 0 | 0 | 0 | 0 | 2 | 0 | 0 | 2 | 0 | 0 | 1 | 0 | 0 | 0 | 19 |
| Jill | White | 34 | 5 | 0 | 1 | 0 | 0 | 0 | 0 | 0 | 0 | 0 | 1 | 0 | 0 | 0 | 0 | 0 | 0 | 0 | 1 | 0 | 0 | 0 | 0 | 1 | 0 | 0 | 0 | 0 | 9 |
| Kareem | Black | 33 | 2 | 0 | 1 | 0 | 0 | 0 | 0 | 0 | 3 | 0 | 10 | 0 | 0 | 0 | 0 | 0 | 0 | 0 | 1 | 0 | 0 | 0 | 1 | 0 | 0 | 0 | 0 | 0 | 18 |
| Katelyn | White | 80 | 0 | 0 | 0 | 0 | 0 | 0 | 0 | 0 | 0 | 0 | 0 | 0 | 0 | 0 | 0 | 0 | 0 | 0 | 0 | 0 | 0 | 0 | 0 | 0 | 0 | 0 | 0 | 0 | 0 |
| Katie | White | 50 | 0 | 0 | 0 | 0 | 0 | 0 | 0 | 0 | 0 | 0 | 0 | 0 | 0 | 0 | 0 | 0 | 0 | 0 | 0 | 0 | 0 | 0 | 0 | 0 | 0 | 0 | 0 | 0 | 0 |
| Keisha | Black | 40 | 1 | 0 | 1 | 0 | 0 | 0 | 0 | 0 | 0 | 0 | 8 | 0 | 0 | 0 | 0 | 0 | 0 | 0 | 1 | 0 | 0 | 0 | 1 | 0 | 0 | 0 | 0 | 0 | 12 |
| Kenya | Black | 4 | 0 | 0 | 0 | 0 | 0 | 0 | 0 | 0 | 0 | 0 | 2 | 0 | 0 | 0 | 0 | 0 | 0 | 0 | 0 | 0 | 0 | 0 | 0 | 0 | 0 | 0 | 0 | 0 | 2 |
| Kristen | White | 34 | 0 | 0 | 0 | 0 | 0 | 0 | 0 | 0 | 0 | 0 | 0 | 0 | 0 | 0 | 0 | 0 | 0 | 0 | 0 | 0 | 0 | 0 | 0 | 0 | 0 | 0 | 0 | 0 | 0 |
| Lakisha | Black | 28 | 0 | 0 | 1 | 0 | 0 | 0 | 0 | 0 | 1 | 0 | 7 | 0 | 0 | 0 | 0 | 0 | 0 | 0 | 2 | 0 | 1 | 0 | 0 | 2 | 0 | 0 | 0 | 0 | 14 |
| Latanya | Black | 28 | 2 | 0 | 1 | 0 | 1 | 0 | 0 | 0 | 1 | 1 | 6 | 0 | 0 | 0 | 0 | 0 | 1 | 0 | 0 | 0 | 0 | 1 | 1 | 0 | 0 | 0 | 0 | 0 | 15 |
| Latisha | Black | 28 | 1 | 0 | 0 | 0 | 0 | 0 | 1 | 0 | 3 | 0 | 11 | 0 | 0 | 0 | 0 | 0 | 0 | 0 | 1 | 0 | 1 | 0 | 0 | 0 | 0 | 0 | 0 | 0 | 18 |
| Latonya | Black | 36 | 0 | 0 | 1 | 0 | 0 | 0 | 0 | 0 | 2 | 1 | 4 | 0 | 0 | 0 | 0 | 1 | 0 | 0 | 3 | 0 | 0 | 1 | 0 | 0 | 0 | 0 | 1 | 0 | 14 |
| Latoya | Black | 27 | 0 | 0 | 2 | 0 | 0 | 0 | 0 | 0 | 0 | 1 | 10 | 0 | 0 | 0 | 0 | 1 | 0 | 0 | 0 | 0 | 1 | 0 | 0 | 0 | 0 | 0 | 0 | 0 | 15 |
| Laurie | White | 28 | 0 | 0 | 0 | 0 | 0 | 0 | 0 | 0 | 1 | 0 | 0 | 0 | 0 | 0 | 0 | 0 | 0 | 0 | 0 | 0 | 0 | 0 | 0 | 0 | 0 | 0 | 0 | 0 | 1 |
| Leroy | Black | 25 | 2 | 0 | 2 | 0 | 1 | 0 | 0 | 0 | 0 | 0 | 9 | 1 | 0 | 0 | 0 | 0 | 0 | 0 | 1 | 1 | 0 | 0 | 0 | 1 | 0 | 1 | 0 | 0 | 19 |
| Luke | White | 60 | 0 | 0 | 0 | 0 | 0 | 0 | 0 | 0 | 0 | 0 | 0 | 0 | 0 | 0 | 0 | 0 | 0 | 0 | 0 | 0 | 0 | 0 | 0 | 0 | 0 | 0 | 0 | 0 | 1 |
| Madeline | White | 66 | 0 | 0 | 0 | 0 | 0 | 0 | 0 | 0 | 0 | 0 | 0 | 0 | 0 | 0 | 0 | 0 | 0 | 0 | 0 | 0 | 0 | 0 | 0 | 0 | 0 | 0 | 0 | 0 | 0 |
| Malik | Black | 18 | 0 | 0 | 1 | 0 | 0 | 0 | 0 | 0 | 0 | 0 | 5 | 0 | 0 | 0 | 0 | 0 | 0 | 0 | 0 | 0 | 0 | 0 | 0 | 1 | 0 | 0 | 0 | 0 | 7 |
| Marquis | Black | 19 | 0 | 0 | 0 | 0 | 0 | 0 | 0 | 0 | 1 | 0 | 6 | 0 | 0 | 0 | 0 | 0 | 0 | 0 | 0 | 0 | 1 | 0 | 0 | 2 | 1 | 0 | 0 | 1 | 12 |
| Matthew | White | 44 | 1 | 0 | 0 | 0 | 0 | 0 | 0 | 0 | 1 | 0 | 4 | 0 | 0 | 0 | 0 | 0 | 0 | 0 | 1 | 0 | 0 | 0 | 0 | 0 | 1 | 0 | 0 | 0 | 8 |
| Meredith | White | 33 | 0 | 0 | 0 | 0 | 0 | 0 | 0 | 0 | 0 | 0 | 2 | 0 | 0 | 0 | 0 | 0 | 0 | 0 | 0 | 0 | 0 | 0 | 0 | 0 | 0 | 0 | 0 | 0 | 2 |
| Molly | White | 70 | 0 | 0 | 0 | 0 | 0 | 0 | 0 | 0 | 0 | 0 | 0 | 0 | 0 | 0 | 0 | 0 | 0 | 0 | 0 | 0 | 0 | 0 | 0 | 0 | 0 | 0 | 0 | 0 | 0 |
| Neil | White | 30 | 1 | 0 | 0 | 0 | 0 | 0 | 0 | 0 | 1 | 1 | 0 | 0 | 0 | 0 | 0 | 0 | 0 | 0 | 0 | 0 | 0 | 0 | 0 | 1 | 0 | 0 | 0 | 0 | 4 |
| Nia | Black | 11 | 0 | 0 | 0 | 0 | 0 | 0 | 0 | 0 | 1 | 0 | 1 | 0 | 0 | 0 | 0 | 0 | 0 | 0 | 0 | 0 | 0 | 0 | 0 | 0 | 0 | 0 | 0 | 0 | 2 |
| Precious | Black | 12 | 0 | 0 | 1 | 0 | 0 | 0 | 0 | 0 | 0 | 0 | 1 | 0 | 0 | 0 | 0 | 0 | 0 | 0 | 0 | 0 | 0 | 0 | 0 | 0 | 0 | 0 | 0 | 0 | 2 |
| Rasheed | Black | 17 | 1 | 0 | 0 | 0 | 0 | 0 | 0 | 0 | 1 | 0 | 1 | 0 | 0 | 0 | 0 | 0 | 0 | 0 | 0 | 0 | 0 | 0 | 0 | 0 | 0 | 1 | 0 | 0 | 3 |
| Shanice | Black | 14 | 2 | 0 | 0 | 0 | 0 | 0 | 1 | 0 | 1 | 0 | 9 | 0 | 0 | 0 | 0 | 0 | 0 | 0 | 1 | 0 | 0 | 0 | 0 | 0 | 0 | 0 | 0 | 0 | 14 |
| Tamika | Black | 29 | 1 | 0 | 4 | 0 | 0 | 0 | 1 | 0 | 3 | 0 | 5 | 0 | 0 | 0 | 0 | 0 | 0 | 0 | 0 | 0 | 0 | 1 | 1 | 0 | 0 | 0 | 0 | 0 | 16 |
| Tanner | White | 30 | 0 | 0 | 0 | 0 | 0 | 0 | 0 | 0 | 0 | 0 | 1 | 0 | 0 | 0 | 0 | 0 | 0 | 0 | 0 | 0 | 0 | 0 | 0 | 0 | 0 | 0 | 0 | 0 | 1 |
| Terrell | Black | 28 | 0 | 0 | 0 | 0 | 0 | 0 | 0 | 0 | 4 | 1 | 9 | 0 | 0 | 0 | 0 | 0 | 0 | 0 | 0 | 0 | 0 | 0 | 0 | 0 | 1 | 0 | 0 | 0 | 15 |
| Tremayne | Black | 27 | 2 | 0 | 0 | 0 | 0 | 0 | 0 | 0 | 1 | 0 | 2 | 1 | 0 | 0 | 0 | 0 | 0 | 0 | 0 | 0 | 0 | 0 | 0 | 0 | 0 | 0 | 0 | 0 | 6 |
| Trevon | Black | 26 | 0 | 0 | 0 | 0 | 0 | 0 | 0 | 0 | 1 | 0 | 2 | 0 | 0 | 0 | 0 | 0 | 0 | 0 | 0 | 0 | 1 | 0 | 0 | 1 | 0 | 0 | 0 | 0 | 5 |
| Tyrone | Black | 36 | 0 | 0 | 1 | 0 | 0 | 0 | 0 | 0 | 8 | 0 | 16 | 0 | 0 | 0 | 0 | 0 | 0 | 0 | 3 | 0 | 0 | 0 | 0 | 2 | 0 | 0 | 0 | 0 | 30 |
| Wyatt | White | 30 | 0 | 0 | 0 | 0 | 0 | 0 | 0 | 0 | 0 | 0 | 0 | 0 | 0 | 0 | 0 | 0 | 0 | 0 | 0 | 0 | 1 | 0 | 0 | 0 | 0 | 0 | 0 | 0 | 1 |
| Totals | | | 33 | 0 | 24 | 0 | 2 | 1 | 6 | 0 | 52 | 7 | 200 | 10 | 0 | 0 | 0 | 6 | 2 | 0 | 30 | 2 | 9 | 2 | 9 | 26 | 3 | 6 | 1 | 1 | 432 |

432

**Figure 21. Distribution of ad templates in Figure 20 by first name as they appeared on Google.com.**

**INSTANT CHECKMATE ADS ON GOOGLE**

| Name | Race | Total Ads | Arrest Ads | | Neutral Ads | |
|---|---|---|---|---|---|---|
| Aaliyah | Black | 0 | 0 | | 0 | |
| Aisha | Black | 23 | 19 | 83% | 4 | 17% |
| Allison | White | 0 | 0 | | 0 | |
| Amy | White | 0 | 0 | | 0 | |
| Anne | White | 0 | 0 | | 0 | |
| Brad | White | 8 | 8 | 100% | 0 | |
| Brendan | White | 11 | 10 | 91% | 1 | 9% |
| Brett | White | 3 | 2 | 67% | 1 | 33% |
| Carrie | White | 2 | 1 | 50% | 1 | 50% |
| Claire | White | 0 | 0 | | 0 | |
| Cody | White | 0 | 0 | | 0 | |
| Connor | White | 0 | 0 | | 0 | |
| Darnell | Black | 18 | 17 | 94% | 1 | 6% |
| DeAndre | Black | 7 | 7 | 100% | 0 | |
| Deja | Black | 10 | 9 | 90% | 1 | 10% |
| DeShawn | Black | 12 | 11 | 92% | 1 | 8% |
| Diamond | Black | | | | 0 | |
| Dustin | White | 1 | 1 | 100% | 0 | |
| Ebony | Black | 26 | 24 | 92% | 2 | 8% |
| Emily | White | 0 | 0 | | 0 | |
| Emma | White | 0 | 0 | | 0 | |
| Geoffrey | White | 5 | 3 | 60% | 2 | 40% |
| Greg | White | 6 | 5 | 83% | 1 | 17% |
| Hakim | Black | 3 | 3 | 100% | 0 | |
| Imani | Black | 3 | 3 | 100% | 0 | |
| Jack | White | 0 | 0 | | 0 | |
| Jake | White | 0 | 0 | | 0 | |
| Jamal | Black | 6 | 5 | 83% | 1 | 17% |
| Jay | White | 4 | 4 | 100% | 0 | |
| Jermaine | Black | 19 | 18 | 95% | 1 | 5% |
| Jill | White | 9 | 4 | 44% | 5 | 56% |
| Kareem | Black | 18 | 16 | 89% | 2 | 11% |
| Katelyn | White | 0 | 0 | | 0 | |
| Katie | White | 0 | 0 | | 0 | |
| Keisha | Black | 12 | 11 | 92% | 1 | 8% |
| Kenya | Black | 2 | 2 | 100% | 0 | |
| Kristen | White | 0 | 0 | | 0 | |
| Lakisha | Black | 14 | 14 | 100% | 0 | |
| Latanya | Black | 15 | 11 | 73% | 4 | 27% |
| Latisha | Black | 18 | 16 | 89% | 2 | 11% |
| Latonya | Black | 14 | 14 | 100% | 0 | |
| Latoya | Black | 15 | 15 | 100% | 0 | |
| Laurie | White | 1 | 1 | 100% | 0 | |
| Leroy | Black | 19 | 16 | 84% | 3 | 16% |
| Luke | White | 0 | 0 | | 0 | |
| Madeline | White | 0 | 0 | | 0 | |
| Malik | Black | 7 | 7 | 100% | 0 | |
| Marquis | Black | 12 | 12 | 100% | 0 | |
| Matthew | White | 8 | 7 | 88% | 1 | 13% |
| Meredith | White | 2 | 2 | 100% | 0 | |
| Molly | White | 0 | 0 | | 0 | |
| Neil | White | 4 | 3 | 75% | 1 | 25% |
| Nia | Black | 2 | 2 | 100% | 0 | |
| Precious | Black | 2 | 2 | 100% | 0 | |
| Rasheed | Black | 3 | 2 | 67% | 1 | 33% |
| Shanice | Black | 14 | 11 | 79% | 3 | 21% |
| Tamika | Black | 16 | 14 | 88% | 2 | 13% |
| Tanner | White | 1 | 1 | 100% | 0 | |
| Terrell | Black | 15 | 15 | 100% | 0 | |
| Tremayne | Black | 6 | 4 | 67% | 2 | 33% |
| Trevon | Black | 5 | 5 | 100% | 0 | |
| Tyrone | Black | 30 | 30 | 100% | 0 | |
| Wyatt | White | 1 | 1 | 100% | 0 | |
| Totals | | 432 | 388 | 90% | 44 | 10% |

| ARREST ADS SORT | NEUTRAL ADS SORT |
|---|---|
| 100% Brad | 56% Jill |
| 100% DeAndre | 50% Carrie |
| 100% Dustin | 40% Geoffrey |
| 100% Hakim | 33% Brett |
| 100% Imani | 33% Rasheed |
| 100% Jay | 33% Tremayne |
| 100% Kenya | 27% Latanya |
| 100% Lakisha | 25% Neil |
| 100% Latonya | 21% Shanice |
| 100% Latoya | 17% Aisha |
| 100% Laurie | 17% Greg |
| 100% Malik | 17% Jamal |
| 100% Marquis | 16% Leroy |
| 100% Meredith | 13% Matthew |
| 100% Nia | 13% Tamika |
| 100% Precious | 11% Kareem |
| 100% Tanner | 11% Latisha |
| 100% Terrell | 10% Deja |
| 100% Trevon | 9% Brendan |
| 100% Tyrone | 8% DeShawn |
| 100% Wyatt | 8% Keisha |
| 95% Jermaine | 8% Ebony |
| 94% Darnell | 6% Darnell |
| 92% Ebony | 5% Jermaine |
| 92% DeShawn | Aaliyah |
| 92% Keisha | Allison |
| 91% Brendan | Amy |
| 90% Deja | Anne |
| 89% Kareem | Brad |
| 89% Latisha | Claire |
| 88% Matthew | Cody |
| 88% Tamika | Connor |
| 84% Leroy | DeAndre |
| 83% Greg | Diamond |
| 83% Jamal | Dustin |
| 83% Aisha | Emily |
| 79% Shanice | Emma |
| 75% Neil | Hakim |
| 73% Latanya | Imani |
| 67% Brett | Jack |
| 67% Rasheed | Jake |
| 67% Tremayne | Jay |
| 60% Geoffrey | Katelyn |
| 50% Carrie | Katie |
| 44% Jill | Kenya |
| Aaliyah | Kristen |
| Allison | Lakisha |
| Amy | Latonya |
| Anne | Latoya |
| Claire | Laurie |
| Cody | Luke |
| Connor | Madeline |
| Diamond | Malik |
| Emily | Marquis |
| Emma | Meredith |
| Jack | Molly |
| Jake | Nia |
| Katelyn | Precious |
| Katie | Tanner |
| Kristen | Terrell |
| Luke | Trevon |
| Madeline | Tyrone |
| Molly | Wyatt |

**Figure 22. Distributions of Instant Checkmate ads having the word "arrest" or not ("neutral") appearing on Google.com.**

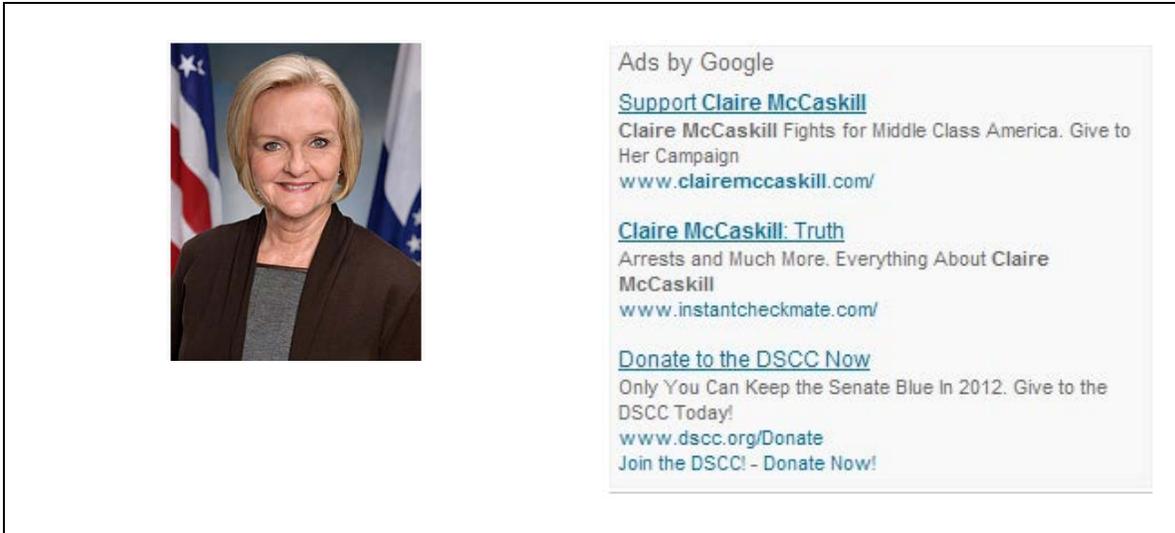

**Figure 23. Example ads displayed in response to search of "*Claire McCaskill*" on Reuters.com (right), Claire McCaskill, U.S. Senator from Missouri (left). An ad having the word "arrest" appears below an ad for her U.S. Senate campaign.**

| | |
|---|---|
| Ads by Google | Ads by Google |
| **Latonya Evans, Arrested?** | **Latisha Smith Located** |
| 1) Enter Name and State. 2) Access Full Background Checks Instantly. | Background Check, Arrest Records, Phone, & Address. Instant, Accurate. |
| www.instantcheckmate.com/ | www.instantcheckmate.com/ |
| Ads by Google | Ads by Google |
| **Latonya Evans's Records** | **Latisha smith: Truth** |
| 1) Enter Name and State. 2) Access Full Background Checks Instantly. | Arrests and Much More. Everything About **Latisha smith** |
| www.instantcheckmate.com/ | www.instantcheckmate.com/ |
| Ad related to **Latonya Evans** ⓘ | Ads related to **Latisha Smith** ⓘ |
| **Latonya Evans's** Records | **Latisha Smith, Arrested?** |
| www.instantcheckmate.com/ | www.instantcheckmate.com/ |
| Did you know **Latonya Evans's** criminal history is searchable? | 1) Enter Name and State. 2) Access Full Background Checks Instantly. |
| | Ad related to **Latisha Smith** ⓘ |
| | **Latisha Smith**, Arrested? - 1) Enter Name and State. |
| | www.instantcheckmate.com/ |
| | 2) Access Full Background Checks Instantly. |

**Figure 24. Examples of different ad copy appearing for searches of "*Latonya Evans*" (left) and "*Latisha Smith*" (right).**

| OVERALL | | | | |
|---|---|---|---|---|
| PeekYou Score | Number of Names with Score | % | Number of Ads | % |
| 2 | 33 | 3% | 11 | 33% |
| 3 | 94 | 8% | 45 | 48% |
| 4 | 170 | 15% | 80 | 47% |
| 5 | 226 | 20% | 111 | 49% |
| 6 | 152 | 13% | 66 | 43% |
| 7 | 382 | 33% | 164 | 43% |
| 8 | 70 | 6% | 21 | 30% |
| 9 | 12 | 1% | 2 | 17% |
| 10 | 4 | 0% | 0 | 0% |
| | | | | |
| Total | 1143 | | 500 | |
| | | | 44% | |
| Average | 5.6 | | | |
| Standard Dev | 1.6 | | | |
| Median | 6 | | | |

| BLACK IDENTIFYING FIRST NAMES | | | | |
|---|---|---|---|---|
| PeekYou Score | Number of Names with Score | % | Number of Ads | % |
| 2 | 33 | 7% | 11 | 33% |
| 3 | 94 | 20% | 45 | 48% |
| 4 | 116 | 24% | 61 | 53% |
| 5 | 158 | 33% | 94 | 59% |
| 6 | 40 | 8% | 26 | 65% |
| 7 | 31 | 6% | 15 | 48% |
| 8 | 5 | 1% | 1 | 20% |
| 9 | 0 | 0% | 0 | |
| 10 | 0 | 0% | 0 | |
| | | | | |
| Total | 477 | | 253 | |
| | 42% | | 53% | |
| Average | 4.4 | | | |
| Standard Dev | 1.3 | | | |
| Median | 4 | | | |

| WHITE IDENTIFYING FIRST NAMES | | | | |
|---|---|---|---|---|
| PeekYou Score | Number of Names with Score | % | Number of Ads | % |
| 2 | 0 | 0% | 0 | |
| 3 | 0 | 0% | 0 | |
| 4 | 54 | 8% | 19 | 35% |
| 5 | 68 | 10% | 17 | 25% |
| 6 | 112 | 17% | 40 | 36% |
| 7 | 351 | 53% | 149 | 42% |
| 8 | 65 | 10% | 20 | 31% |
| 9 | 12 | 2% | 2 | 17% |
| 10 | 4 | 1% | 0 | 0% |
| | | | | |
| Total | 666 | | 247 | |
| | 58% | | 37% | |
| Average | 6.5 | | | |
| Standard Dev | 1.1 | | | |
| Median | 7 | | | |

**Figure 25. Distributions of Netizen names and ad delivery by PeekYou scores for those names having PeekYou scores, which are values PeekYou assigns to names as an estimate of the person's presence on the Web.**

## Conclusion and Future Work

This study raises more questions than it answers.  Here is the one answer provided. Our hypothesis states that no difference exists in the delivery of ads suggestive of an arrest record based on searches of racially associated names. Our findings reject this hypothesis. A greater percentage of ads having "arrest" in ad text appeared for black identifying first names than for white identifying first names in searches on Reuters.com, on Google.com, and in subsets of the sample.  Results of Chi-Square tests on these patterns were statistically significant.  On Reuters.com, a host of Google AdSense ads, a black-identifying name was 25% more likely to get an ad suggestive of an arrest record, $X^2(1)=14.32$, $p < 0.001$; there is less than a 0.1% probability that these data can be explained by chance.

*Why is this discrimination occurring*?  Is this Instant Checkmate, Google, or society's fault? Answering these questions is beyond the scope of this writing, but navigating the terrain requires further information about the inner workings of Google AdSense.  Google understands that an advertiser may not know which ad copy will work best, so an advertiser may give multiple templates for the same search string and the "Google algorithm" learns over time which ad text gets the most clicks from viewers of the ad.  It does this by assigning weights (or probabilities) based on the click history of each ad copy.  At first all possible ad copies are weighted the same, they are all equally likely to produce a click.  Over time, as people tend to click one version of ad text over others, the weights change, so the ad text getting the most clicks eventually displays more frequently. This approach aligns the financial interests of Google, as the ad deliverer, with the advertiser.  Figure 24 provides examples in which Instant Checkmate provided multiple ad templates for searches of "*Latonya Evans*" and "*Latisha Smith*".

Did Instant Checkmate provide ad templates suggestive of arrest disproportionately to black-identifying names?[3] Or, did Instant Checkmate provide roughly the same templates evenly across racially associated names but society clicked ads suggestive of arrest more often for black identifying names? Google uses cloud-caching strategies to deliver ads quickly, might these strategies bias ad delivery towards ad templates previously loaded in the cloud cache? Is there a combinatorial effect?

This paper is a start and more research is needed; however, online advertising is dynamic and easy to change.  In order to preserve research opportunities, prior to any announcement of this work, I captured additional results for 50 hits on 2184 names across 30 websites serving Google Ads to learn the underlying distributions of ad occurrences per name. While analyzing these data may prove illuminating, in the end, the basic message presented in this writing does not change. There is discrimination in delivery of these ads.

---

[3] During a conference call with the founders of Instant Checkmate and their lawyer on December 21, 2012, the company's representatives asserted that Instant Checkmate gave the same ad text to Google for groups of last names (not first names).

In the broader picture, technology can do more to thwart discriminatory effects and harmonize with societal norms. Ads responding to name searches appear in a specific information context and technology controls that context. A reader enters a name then views web content and news stories specific to that name. Dynamic ads are a part of that context. Alongside news stories about high school athletes and children can be ads bearing the child's name and suggesting arrest. This seems concerning on many levels. For example, even if the child has an arrest record, juvenile records are typically exempt from public record disclosure. The juxtaposition of ads also provide context. Claire McCaskill provides an example where an ad suggestive of arrest appears alongside an ad for her U.S. Senate campaign. Search and ad technology already reason extensively about context and appropriateness when deciding the best content to deliver to the reader [13]. Many factors are often known about the reader at the time of ad delivery, e.g., browsing history, geographical location, and shopping behavior [14]. With some expansion, technology could additionally reason about social and legal implications of content and context too. For example, well-known computer scientist Cynthia Dwork and her colleagues have already been working on algorithms that assure racial fairness [15]. This area seems ripe for further research and development.

## Acknowledgements

The author thanks Ben Edelman, Claudine Gay, Gary King, Annie Lewis, and weekly Topics in Privacy participants (David Abrams, Micah Altman, Merce Crosas, Bob Gelman, Harry Lewis, Joe Pato, and Salil Vadhan) for discussions, the Institute for Quantitative Social Science, the Department of Government, Dean Smith, Jim Waldo and my students at Harvard for an awesome work environment, Adam Tanner for first suspecting a pattern, and Diane Lopez and Matthew Fox in Harvard's Office of the General Counsel for making publication possible in the face of legal threats. Data from this study are available at foreverdata.org and the IQSS Dataverse Network. Supported in part by NSF grant CNS-1237235 and a gift from Google, Inc.